\documentclass[letterpaper]{article}
\usepackage{aaai}
\usepackage{times}
\usepackage{helvet}
\usepackage{courier}
\usepackage{subfigure}
\usepackage{xspace}
\usepackage{amssymb, amsmath}
\usepackage{comment}
\usepackage{color}
\usepackage{colortbl}
\usepackage{graphicx}
\newcommand{\checkins}{check-ins\xspace}
\newcommand{\checkin}{check-in\xspace}

\begin{document}
% The file aaai.sty is the style file for AAAI Press
% proceedings, working notes, and technical reports.
%
\title{You are What you Eat (and Drink): \\Identifying Cultural Boundaries by Analyzing Food \& Drink Habits in Foursquare}

\author{Thiago H Silva$^\star$,
  Pedro O S Vaz de Melo$^\star$,
  Jussara Almeida$^\star$,
  Mirco Musolesi$^\dag$,
  Antonio Loureiro$^\star$\\  
  $^\star$Department of Computer Science, Universidade Federal de Minas Gerais, Belo Horizonte, MG, Brazil\\
  $^{\dag}$School of Computer Science, University of Birmingham, Birmingham, UK\\
  \{thiagohs, olmo, jussara, loureiro\}@dcc.ufmg.br, m.musolesi@cs.bham.ac.uk
}

\maketitle
\begin{abstract}
Food and drink are two of the most basic needs of human beings. However, as society evolved, food and drink became also a strong cultural aspect, being able to describe strong differences among people. Traditional methods used to analyze cross-cultural differences are mainly based on surveys and, for this reason, they are very difficult to represent a significant statistical sample at a global scale. In this paper, we propose a new methodology to identify cultural boundaries and similarities across populations at different scales based on the analysis of Foursquare check-ins. This approach might be useful not only for economic purposes, but also to support existing and novel marketing and social applications. Our methodology consists of the following steps. First, we map food and drink related check-ins extracted from Foursquare into users' cultural preferences. Second, we identify particular individual preferences, such as the taste for a certain type of food or drink, e.g., pizza or sake, as well as 
temporal habits, such as the time and day of the week when an individual goes to a restaurant or a bar. Third, we show how to analyze this information to assess the cultural distance between two countries, cities or even areas of a city. Fourth, we apply a simple clustering technique, using this cultural distance measure, to draw cultural boundaries across countries, cities and regions.
\end{abstract}

\section{Introduction}

What are your eating and drinking habits? How different are they from a typical individual from Japan or Germany? It is impossible to answer these questions without addressing the cultural features within groups of individuals. However, culture is such a complex and interesting concept that no simple definition or measurement can capture it. Among the various aspects that define the culture of a society (or person), one may cite its arts, religious beliefs, literature, manners and scholarly pursuits. Moreover, as Counihan~\cite{foodCulture:Counihan}, and Cochrane and Bal~\cite{drinkBehav} pointed out, eating and drinking habits are also fundamental elements in a culture and may significantly mark social differences, boundaries, bonds, and contradictions. Since eating and drinking habits have such importance for a culture, we here address the topic of investigating and analyzing life and idiosyncrasies of different societies through them.

How can we analyze eating and drinking habits at a large scale? Nowadays, the study of social behavior at a large scale is possible thanks to the increasing popularity of smart phones and location sharing systems such as Foursquare. By means of these technologies, it is possible to sense human activities related to food and drink practices (e.g., restaurant visiting patterns) in large geographical areas, such as cities or entire countries. Foursquare, created in 2009, registered 5 million users in December 2010 and 45 million users in January 2014. Data generated by this popular application triggers unprecedented opportunities to measure cultural differences at a global scale and at low cost~\cite{silvaISCC2013}.

In this work, we propose a new methodology for identifying cultural boundaries and similarities across populations using self-reported cultural preferences recorded in location-based social networks (LBSNs). Our methodology, which is here demonstrated using data collected from Foursquare, consists of the following steps. First, we map food and drink \checkins extracted from Foursquare into users' cultural preferences. By exploring this mapping, we are able to identify particular individual preferences, such as the taste for barbecue or sake. Food and drink individual preferences, as shown in this paper, are good indicators of cultural similarities between users. We then show how to extract features from Foursquare data that are able to delineate and describe regions that have common cultural elements, defining signatures that represent cultural differences between distinct areas around the planet. To that end, we investigate two properties of food and drink preferences: geographical and temporal 
characteristics. Next, we apply a simple clustering technique, namely $k$-means, to show the ``cultural distance'' between two countries, cities or even regions of a city, allowing us to draw cultural boundaries across them.

Unlike previous efforts, which used survey data, our work is based on a dynamic and publicly available Web dataset representing habits of a much larger and diverse population. Besides being globally scalable, our methodology also allows the identification of cultural dynamics more quickly than traditional methods (e.g., surveys), since one may observe how countries or cities are becoming more culturally similar or distinct over time.

The correct identification of cultural boundaries is useful in many fields and applications. Rather than using traditional methods to identify cultural differences, the proposed method is an easier and cheaper way to perform this task across many regions of the world, because it is based on data voluntarily shared by users on Web services. Moreover, since culture is an important aspect for economic reasons~\cite{garcia2013}, our methodology is valuable for companies that have businesses in one country and want to verify the compatibility of preferences across different markets. Another application that could rely on our methodology is a place recommendation system, which is useful for visitors and residents of a city. Foursquare estimates that only 10\% to 15\% of searches on Foursquare are for specific places~\cite{newsFoursquareReco}. Much more often users are searching within broader categories, such as ``sushi''~\cite{newsFoursquareReco}. Based on this information, systems like Foursquare and 
other location-based search engines, as the one proposed in \cite{shankar:crowds}, could benefit from the introduction of new criteria and mechanisms in their recommendation systems that consider cultural differences between areas. For instance, a person who enjoyed a specific area of Manhattan could receive a recommendation of a similar area when visiting London.

The rest of this paper is organized as follows. Section~\ref{relatedwork} presents the related work. Section~\ref{sec:extract} describes our dataset and the core of our methodology for extracting cultural preferences from location-based social networks. Section~\ref{secIndividualPrefs} investigates the cultural similarities between individuals, and shows that food and drink \checkins outperforms \checkins given in all types of places in this case. Section~\ref{sec:featuresExtract} shows how to extract cultural signatures for different areas of the globe and explore the similarities among them, while Section~\ref{secCulturalBound} applies this knowledge to analyze the implicit cultural boundaries that exist for different cultural aspects of the society. Finally, Section \ref{sec:conclusion} summarizes our contributions and discusses some possibilities of future work.

\section{Related Work}\label{relatedwork}

Several studies have focused on the spatial properties of data shared in location-based services such as Foursquare \cite{Scellato:2011:SocioSpacialProp,Cho:2011:friendshipMobility,NoulasSMP11:Pat4square}. However, those prior efforts aimed at investigating user mobility patterns or social network properties and their implications. More recently, researchers have started looking at user activity as another data source that can be leveraged for studying social interactions~\cite{Sakaki:2010}. Based on this principle, there have been many studies to extract new insights about city dynamics such as, for example, their key characteristics and the behavior of their citizens. For instance, Cranshaw et {al.}~\cite{cranshaw:livehoods} presented a model to extract distinct regions of a city according to current collective activity patterns. Similarly, Noulas et {al.}~\cite{Noulas2011} proposed an approach to classify areas of a city by using all venues' categories of Foursquare.

Some recent studies have shown how the use of Web systems vary across countries. For example, Hochman et {al.}~\cite{Hochman:2012} investigated color preferences in pictures shared through Instagram, showing considerable differences in the preferences across countries with distinct cultures. Garcia-Gavilanes et {al.}~\cite{garcia2013} and Poblete et {al.}~\cite{poblete2011all} studied variations of Twitter usage across countries. In particular, Garcia-Gavilanes et {al.} showed that cultural differences are not only visible in the real world but also observed on Twitter.

Cross-cultural studies (i.e., the study of cultural differences) do not constitute a new research area. Indeed, they have been carried out by researchers working in the social sciences, particularly in cultural anthropology and psychology~\cite{murdock1949social}. Despite globalization and many other technological revolutions~\cite{Blossfeld2005}, group formation might lead to the emergence of cultural boundaries that exist for millennia across populations~\cite{barth1969ethnic}. Axelrod~\cite{Axelrod:1997p5522} proposed a model to explain the formation and persistence of these cultural boundaries, which are basically a consequence of two key phenomena: social influence~\cite{festinger1967social} and homophily~\cite{mcpherson2001birds}. While homophily dictates that only culturally similar individuals are likely to interact, social influence makes individuals more similar as they interact. In a long run, these two phenomena lead to very culturally distinct groups of individuals, delimited by the so-called \textit{cultural boundaries}.

\section{Extracting Cultural Preferences} \label{sec:extract}

In this section we present our dataset and our methodology for extracting cultural preferences from LBSNs.

\subsection{Mapping User Preferences}\label{sec:MappingPreferences}

One of the biggest challenges in the analysis of cultural differences among people and regions is finding the appropriate empirical data to use. The common approach to overcome this challenge is the use of surveys based on questionnaires filled during face-to-face interviews~\cite{valori:2012}, such as the Eurobarometer dataset~\cite{ebtrend-appendix}. Through these questionnaires, individual preferences, such as the taste for coffee and fast food, can be mapped into multidimensional vectors representing (and characterizing) each interviewee. From these vectors, it is possible, for instance, to quantify how similar or different two individuals are.

Although survey data are broadly used in the analysis of cultures, there are some severe constraints in its use, which are well known to researchers. First, surveys are costly and do not scale up. That is, it is hard to obtain data of millions, or even thousands of people. Second, they provide static information, i.e., they reflect the preferences of users at a specific point in time. If some of the preferences change for a significant amount of the interviewed people, such as the taste for online gaming instead of street ball playing, the data is compromised.

In order to overcome the aforementioned constraints, we propose the use of publicly available data from LBSNs to map individual preferences. LBSNs can be accessed everywhere by anyone who has an Internet connection, solving the scalability problem and allowing data from (potentially) the entire world to be collected~\cite{silvaISCC2013}. Moreover, these systems are dynamic, being able to capture the behavioral changes of their users when they occur, which solve the second mentioned constraint. However, data from such systems can be used if and only if they meet the requirements:

\begin{itemize}
\item \textbf{[R1]} It is possible to associate a user to its location;
\item \textbf{[R2]} It is possible to extract a finite set of preferences from the data that is generated by the system;
\item \textbf{[R3]} It is possible to map users' actions in the system into the preferences defined in \textbf{[R2]}.
\end{itemize}

Considering that these requirements are met, a dataset containing individual activities of $N$ users of a LBSN can be used to map preferences as follows. First, associate each user $n_i$ with a location $l_i$, which may be a country, a city or even a region within a city. Then, define a set of $m$ individual preferences (or features) $f_1, f_2, \ldots, f_m$ that can be extracted from the dataset, which may represent the taste for the most varied things, such as Japanese food or a certain football team. Finally, map the activities of each individual $n_i$ into an $m$-dimensional vector of preferences $F_i = f_{1^i}, f_{2^i}, \ldots, f_{m^i}$ that characterizes the person's tastes, the same type of vector that is usually created from survey data~\cite{valori:2012}.

Since the preference vector $F_i$ is generated from self-reported temporal data of an individual $n_i$, we may populate and modify it in various ways. For instance, we can use a binary representation, where $f_{k^i} = {0|1}$ represents whether user $n_i$ has or not preference $f_k$ (e.g., whether a person likes/dislikes a certain type of food), respectively. Alternatively, we may consider the intensity at which a user likes a feature, inferred from the number of times the corresponding preference is reported in the person's data, i.e., $f_{k^i} = [0;\infty)$. In Section~\ref{secIndividualPrefs}, we adopt a binary representation. Finally, one can group individuals by their geographical regions and sum up their preference vectors to characterize their regions. We adopt this approach in Section~\ref{sec:featuresExtract} to build preference vectors for regions (instead of individuals).

%===============================================================================================================

\subsection{Data Description}\label{sec:generalView}

In this work, the dataset used to infer user preferences was collected from one of the currently most popular location based social networks, namely Foursquare. We collected this data from Twitter\footnote{http://www.twitter.com.}, since Foursquare \checkins are not publicly available by default. Approximately 4.7 million tweets containing \checkins were gathered, each one providing a URL to the Foursquare website where information about the venue, in particular its geographic location and category, was acquired. In the dataset, each \checkin consists of the latitude, longitude, identifier, and category of the venue as well as the time when the \checkin was done. Foursquare venues are grouped into eight categories: Arts \& Entertainment; College \& University; Professional \& Other Places; Residences; Great Outdoors; Shops \& Services; Nightlife Spots; and Food. Each category, in turn, has subcategories. For example, Rock Club and Concert Hall are subcategories of Nightlife Spots. In order to show that our 
methodology is able to capture cultural dynamics in short time windows, we use a dataset that spans a single week of April 2012.

Moreover, since we are primarily interested in food and drink habits, we manually grouped relevant subcategories of the Food and Nightlife Spots categories into three classes: Drink, Fast Food, and Slow Food places. We did this by excluding some subcategories that are not related to these three classes (e.g. Rock Club and Concert Hall) and moving some subcategories (e.g. Coffee Shop and Tea Room) from the Food category to the Drink class. Besides that we also disregard the category Restaurant, because it is a sort of meta category that could fit in any of the two classes of food. After this manual classification process, the Drink class ended up with 279,650 \checkins, 106,152 unique venues and 162,891 unique users; the Fast Food class with 410,592 \checkins, 193,541 unique venues, and 230,846 unique users; and the Slow Food class with 394,042 \checkins, 198,565 unique venues, and 231,651 unique users. Moreover, the Drink class has 21 subcategories (e.g., brewery, karaoke bar, and pub), whereas 
the Fast Food class has 27 subcategories (e.g., bakery, burger joint, and wings joint) and the Slow Food class has 53 subcategories, including Chinese restaurant, Steakhouse, and Greek restaurant.

To provide an idea about the size of the user population \mbox{LBSNs} can reach, consider the World Values Survey\footnote{http://www.worldvaluessurvey.org.} project. That study is maybe the most comprehensive investigation of political and sociocultural change worldwide, which was conducted from 1981 to 2008 in 87 societies, with about 256,000 interviews. Observe that our one-week dataset has a population of users of the same order of magnitude of the number of interviews performed in that project in almost three decades.

\begin{figure}[t!]
\centering
   \subfigure[Drink]
   {
   \includegraphics[width=55mm]{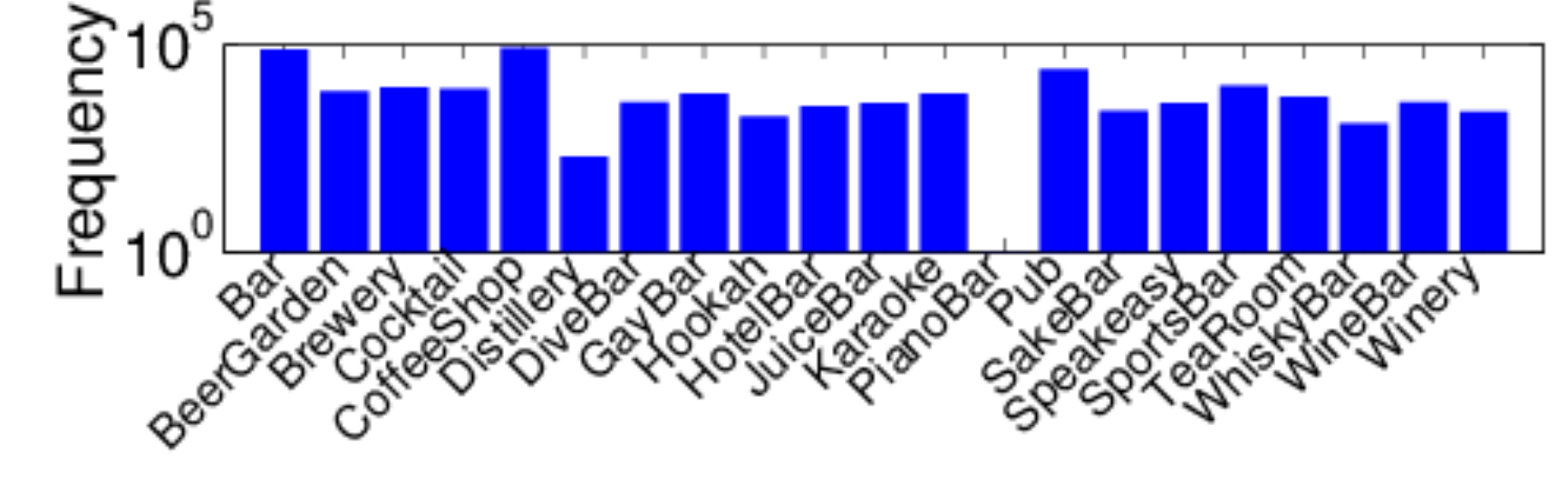}}\vspace{-2mm}
  \subfigure[Fast Food]
    {%\includegraphics[width=65mm,height=32mm]{../images/fastFood/allLocalities/numCheckinsCategory/numChksSubcat_log2.pdf}}\vspace{-4mm}
    \includegraphics[width=65mm]{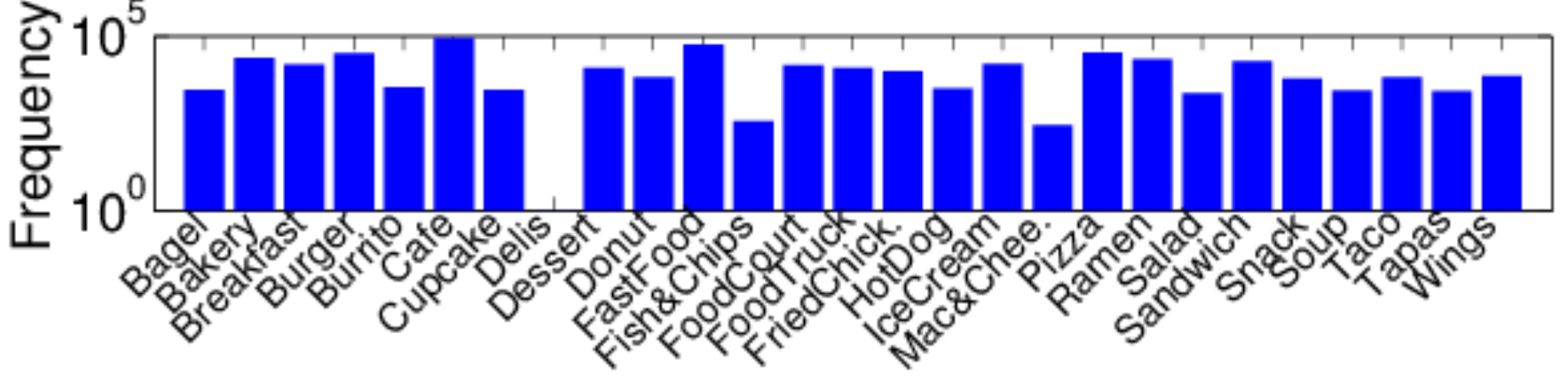}}
 \subfigure[Slow Food]
    {%\includegraphics[width=95mm,height=32mm]{../images/slowFood/allLocalities/numCheckinsCategory/numChksSub_log2.pdf}}\vspace{-4mm}
    \includegraphics[width=85mm]{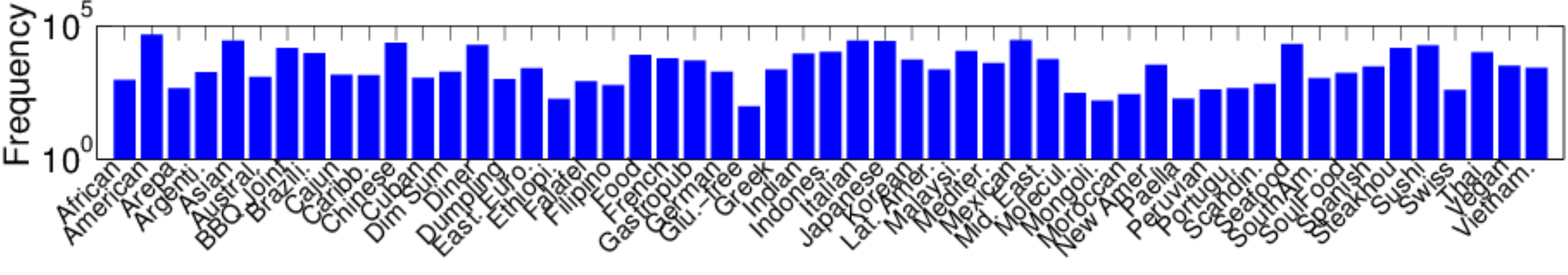}}\vspace{-4mm}
\caption{Frequency of \checkins at all subcategories of the three analyzed classes. The names of some places are abbreviated but the semantics of the names is preserved.}
\label{fig:numChksClasses}
\end{figure}

 %=======================================================================================================================
\subsection{Mapping Foursquare Data into User Preferences}\label{secMappingPreferences}

Several characteristics of human beings are not directly observable, such as personality traits. Thus, we rely on face-to-face interactions or online signals to discover the presence of those hidden qualities~\cite{pentland2010signal}. In this direction, a LBSN \checkin can be considered as a signal because it is a perceivable feature/action that expresses the preference of a user for a certain type of place. With that in mind, we use Foursquare \checkins to represent user preferences regarding food and drink places. Specifically, we use the three main classes defined in Section~\ref{sec:generalView}, namely, \textit{Drink}, \textit{Fast Food}, and \textit{Slow Food}.

Figures~\ref{fig:numChksClasses}a, \ref{fig:numChksClasses}b, and \ref{fig:numChksClasses}c show the frequency of \checkins at each subcategory of the Drink, Fast Food, and Slow Food classes, respectively, so we can have a general idea about the popularity of user preferences for different food and drink related places. These figures show the popularity of different places according to people's preferences worldwide. Note that Coffee Shop and Bar are the two most popular subcategories of Drink places, with 86,310 and 81,124 \checkins, respectively. The two most popular Fast Food subcategories are Caf\'{e}\footnote{Like in many European countries, this term is referred as a restaurant primarily serving coffee as well as pastries.} and Fast Food Restaurant, with 91,303 and 56,648 \checkins, respectively. Finally, American Restaurant (47,373 \checkins), and Mexican Restaurant (28,712 \checkins) are the two most visited subcategories of Slow Food places.

In this dataset, a user is represented by a vector of $m=$101 features corresponding to the 101 subcategories that comprise the three classes we have defined. A feature $f_i \in F = \{f_1, f_2, \ldots,f_{101}\}$ is equal to 1 if a user made at least one \checkin at $f_i$, and 0 otherwise. In this way, a feature vector represents the positive and negative preferences of a user for fast food, slow food and drink subcategories. With that, a finite set of preferences is extracted (requirement \textbf{[R2]}, see definition in Section~\ref{sec:MappingPreferences}) and users' actions are mapped into this set (requirement \textbf{[R3]}). To associate a user with a location (requirement \textbf{[R1]}), we analyzed the GPS coordinates of all \checkins performed by the user. If all \checkins performed are from the same country, according to the free reverse geocoding API offered by Yahoo\footnote{http://developer.yahoo.com.}, we assume that the user taken into consideration is from that country. Otherwise, we do not 
consider the user in our analysis. In this way, we minimize the wrong association of a user with a country. Following this procedure, approximately 1\% of the users were disregarded from our analysis.

\section{Cultural Analysis of Individuals}\label{secIndividualPrefs}

In this section, we use the map of preferences presented in Section~\ref{secMappingPreferences} to analyze the individual preferences of users, showing, among other results, that food and drink preferences are good indicators of cultural similarities.

In order to assess the cultural similarities among users, we construct a similarity network $G_s=(V_s, E_s)$, where $s$ is a similarity threshold used to build the network, vertices $V_s$ represent the set of users, and an edge $(v_i,v_j)$ exists in $E_s$ if users $v_i$ and $v_j$ have a similarity score above $s$. The similarity score $s_{i,j}$ between two users $v_i$ and $v_j$ is the Jaccard index (JI) between their preference vectors\footnote{The Jaccard index of sets A and B is computed as $\frac{A \cap B}{A \cup B}$.} multiplied by 100. In this way, $s_{i,j}$ varies from 0 to 100 and measures the percentage of preferences shared by the users $v_i$ and $v_j$. For example, considering a similarity threshold $s=$ 65 (or 65\%-network\footnote{Network created with a threshold $s$ is referred to as $s$-network.}), there is an edge between vertices $v_1$ and $v_2$ if the corresponding users have, at least, 65\% of preferences in common. We have built two similarities 
networks: $G_{s}^1$; and $G_{s}^2$. The network $G_{s}^1$ considers only food and drink preferences, i.e., only \checkins at food and drink places. On the other hand, $G_{s}^2$ consider all preferences, i.e., all Foursquare subcategories, including food and drink venues. To build both networks we consider only the users who performed at least 7 \checkins in the dataset (i.e., at least one \checkin per day on average). In total, 28,038 users were considered in $G_{s}^1$ and 194,902 in $G_{s}^2$. Moreover, isolated nodes were disregarded. We here consider the following values of $s \in \{$65, 70, 75, 80, 85, 90, 95, 100$\}$. Note that $G_{s}^1$ and $G_{s}^2$ are undirected unweighted and symmetric graphs.

\begin{figure}[t!]
\centering
 % \subfigure[\# of connected components]
  %     {\includegraphics[width=.27\textwidth]{../images/analiseGrafos/numCompConexos.pdf}}
  \subfigure[\% of users in the 2$^{\mbox{\scriptsize nd}}$ largest comp. $G_{s}^1$]
      {\includegraphics[width=.15\textwidth]{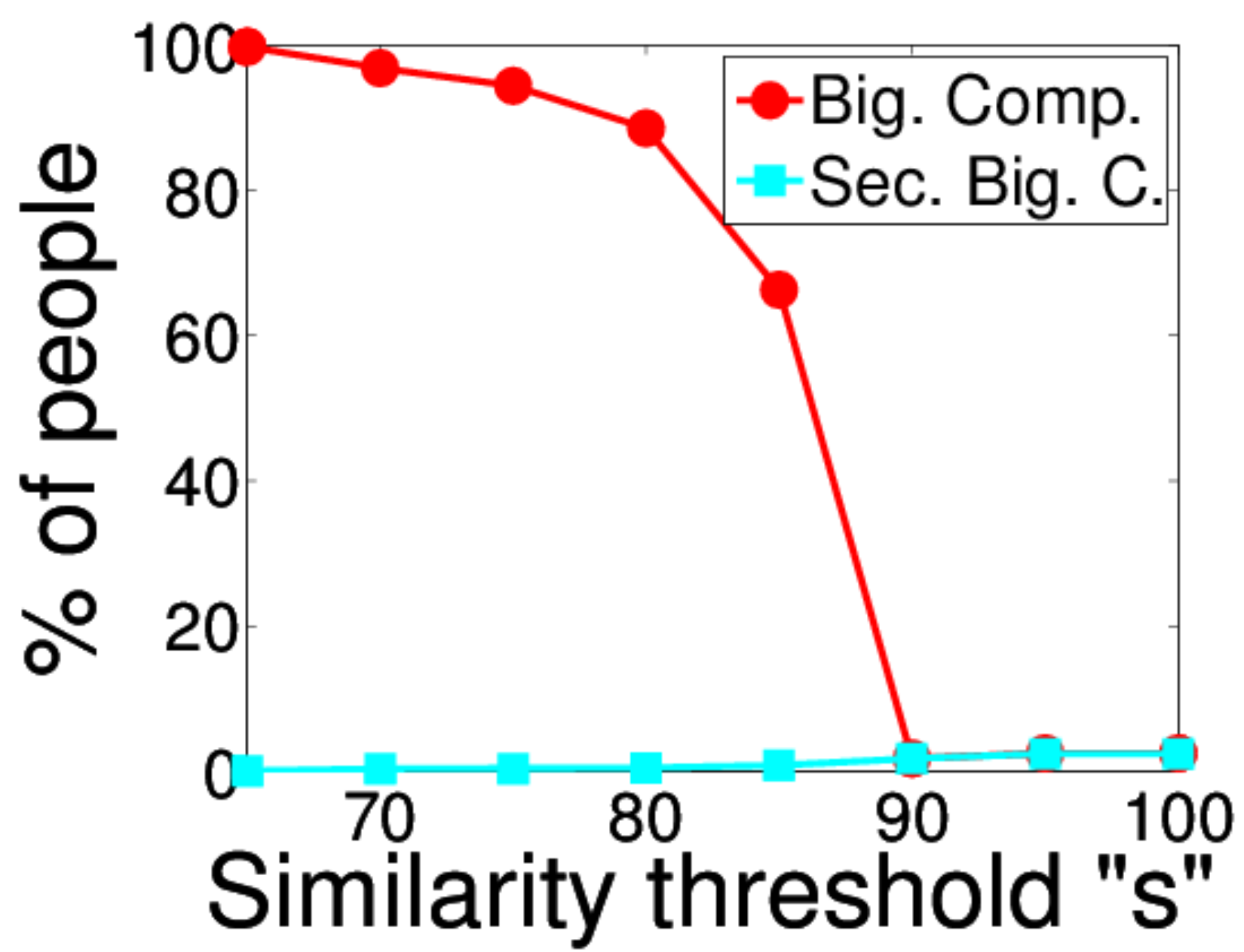}}
   %\subfigure[Average shortest path]
   %     {\includegraphics[width=.225\textwidth]{../images/analiseGrafos/caminhoMinimoMedioMaiorComp.pdf}}
  \subfigure[Assortat. $G_{s}^1$]
      {\includegraphics[width=.15\textwidth]{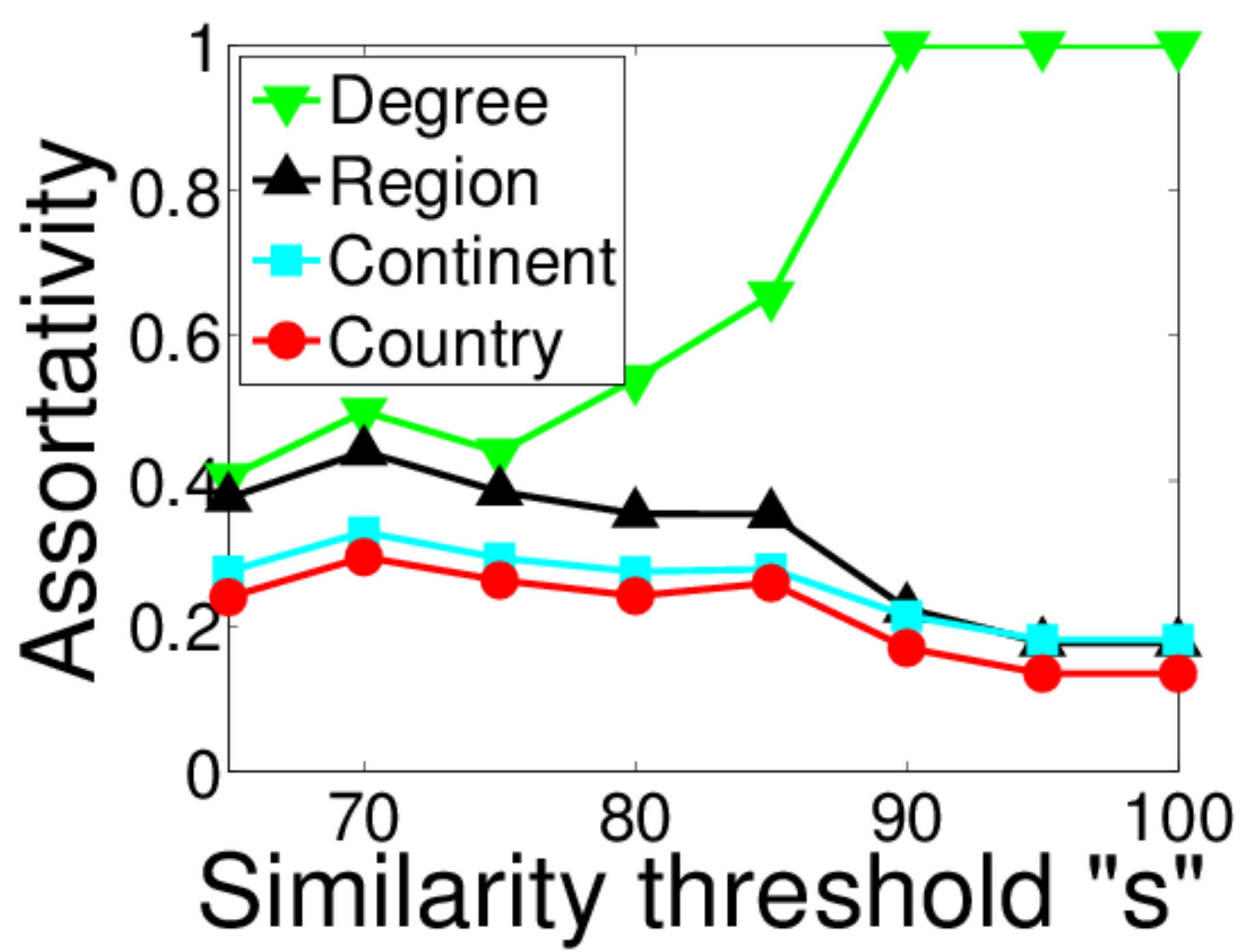}}
  \subfigure[Assortat. $G_{s}^2$]
      {\includegraphics[width=.15\textwidth]{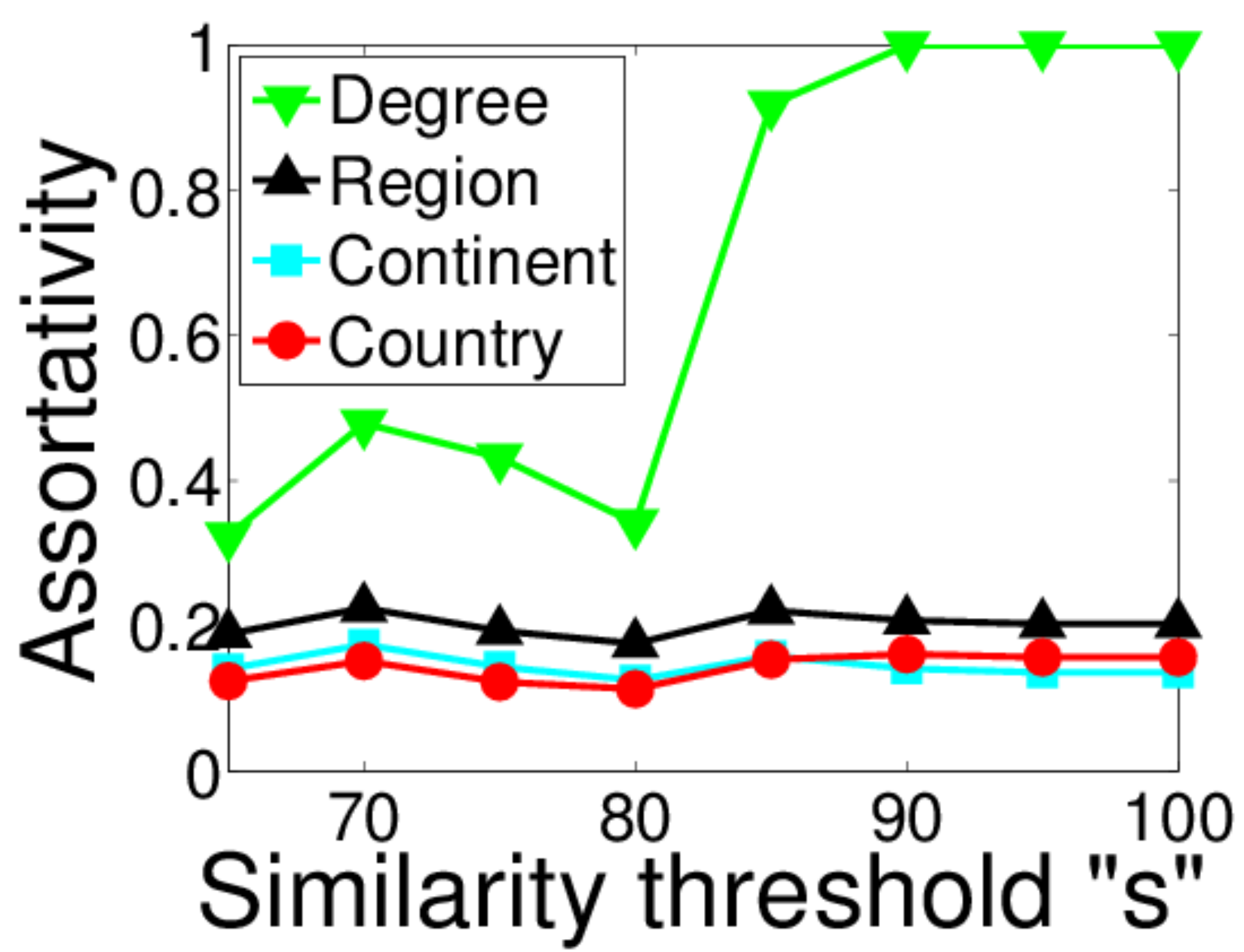}}\vspace{-4mm}
%\caption{General network metrics for all similarity networks considered.}\label{fig:participatorySensorNet}
\caption{General metrics for all similarity networks.}\label{fig:participatorySensorNet}
\end{figure}

We first analyze relevant properties of $G_{s}^1$ and $G_{s}^2$. Figure~\ref{fig:participatorySensorNet}a shows the percentage of vertices (i.e., users) in the two largest components of the network $G_{s}^1$, for various values of $s$ (figure omitted for the network $G_{s}^2$ due to space limitations). Figure~\ref{fig:participatorySensorNet}a shows that the largest component of the 65\%-network practically contains all nodes. The percentage of users in the largest component slowly decreases as the similarity threshold increases, until $s$ reaches 85. For larger values of $s$, the number of users in the largest component drops sharply, becoming comparable to the size of the second largest component. This is explained by observing networks built using large values for $s$, such as the 100\%-network, where every component is composed of very similar users. Since users with very similar preferences are rare, the largest components tend not to have very large differences in size. We note that the results for the 
network $G_{s}^2$ are similar to those observed for the network $G_{s}^1$, for example, the largest component of the 65\%-network also contains practically all nodes.

\begin{figure*}[t!]
\centering
\subfigure[Drink]
{\includegraphics[width=0.325\textwidth]{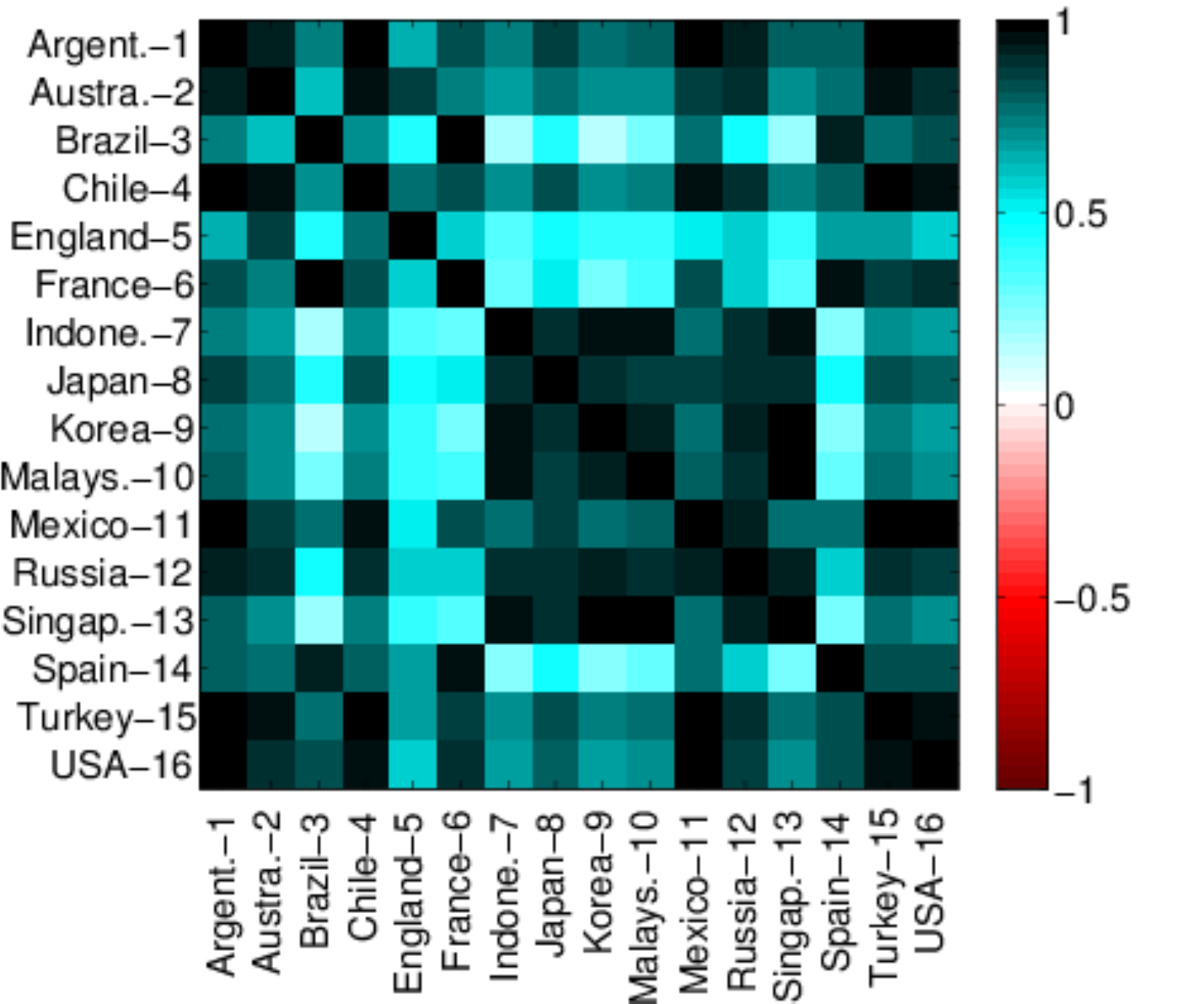}}
\subfigure[Fast Food]
{\includegraphics[width=0.325\textwidth]{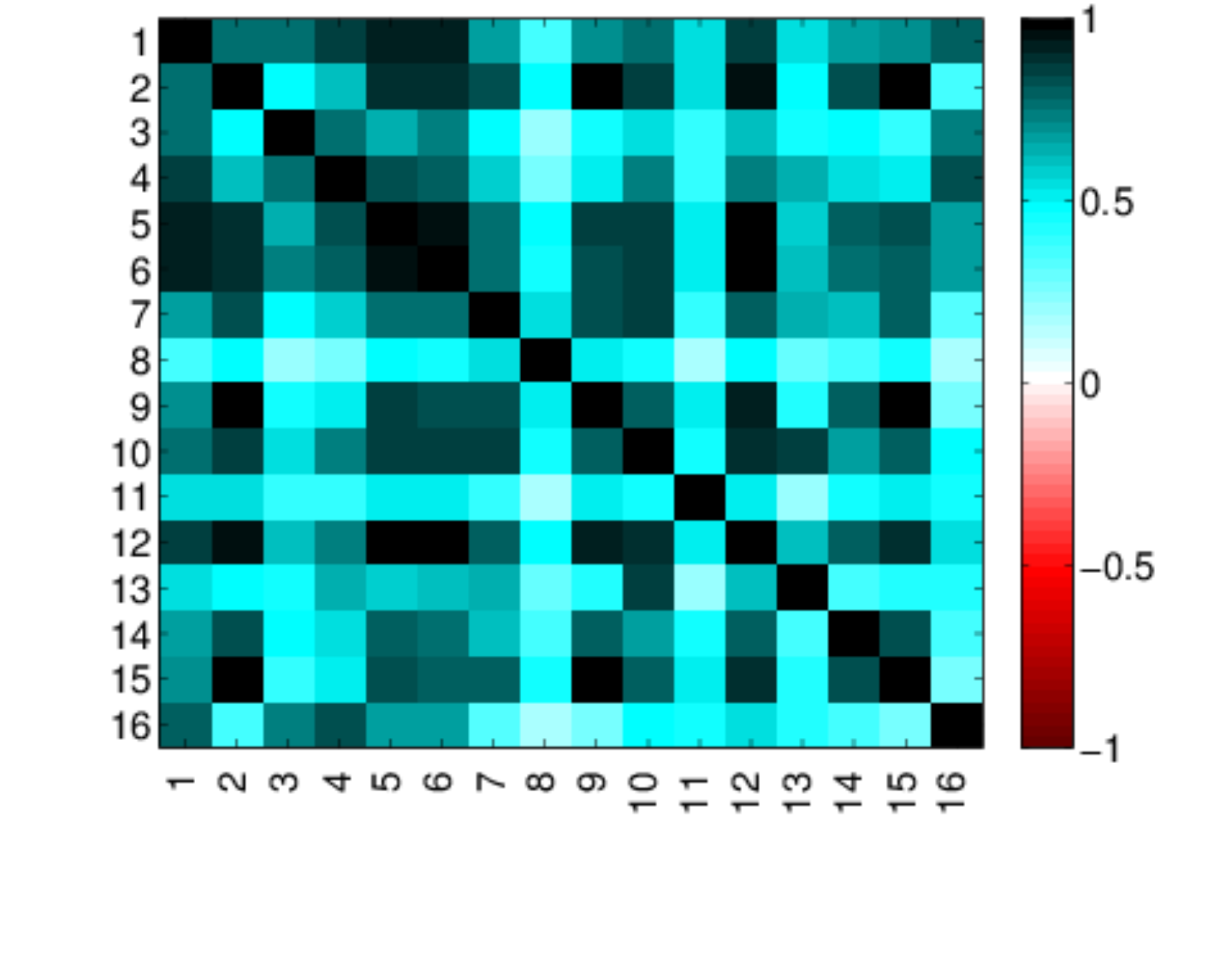}}
\subfigure[Slow Food]
{\includegraphics[width=0.325\textwidth]{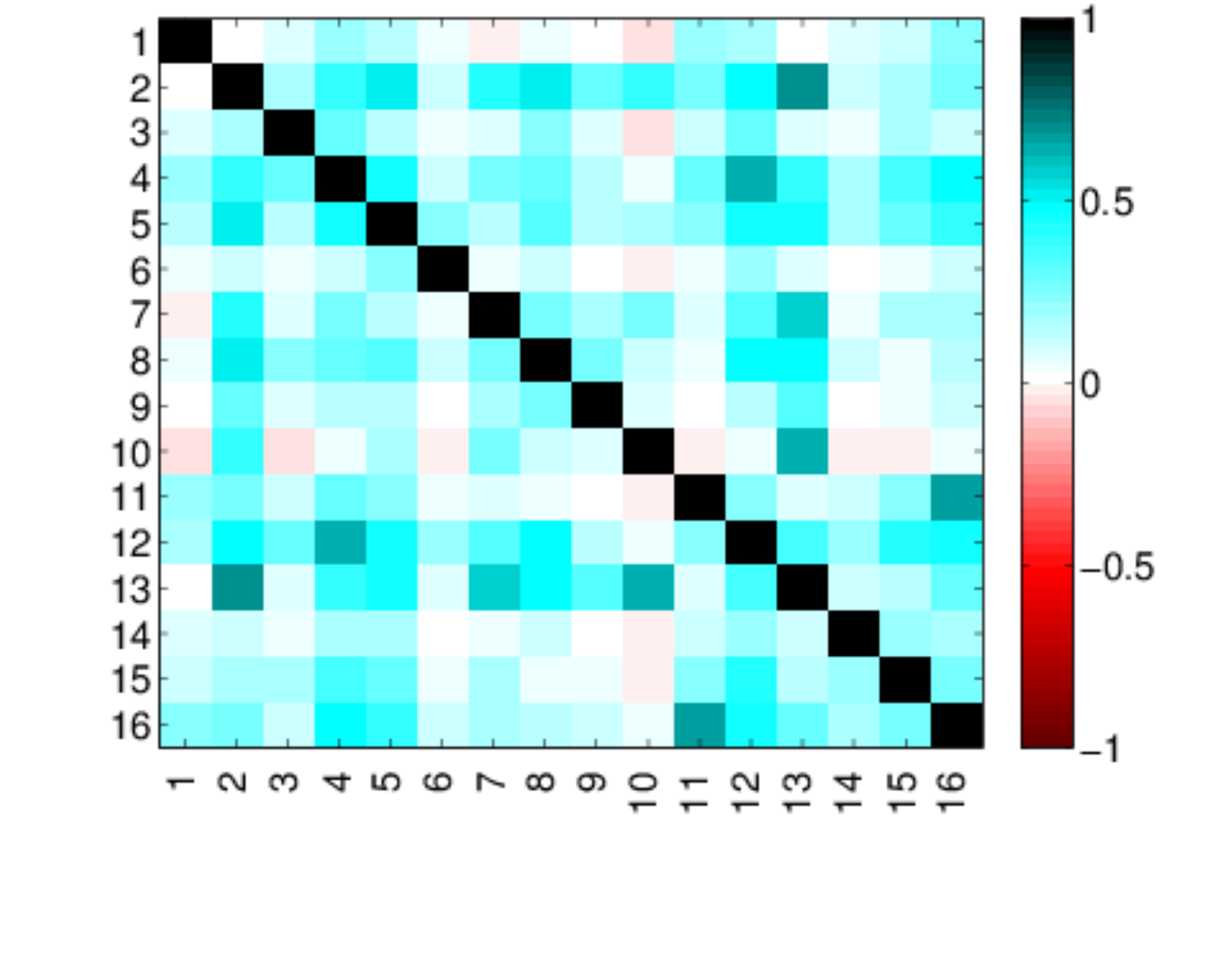}}\vspace{-4mm}
\caption{Correlation of preferences between countries.}\label{fig:correlationCountries}
\end{figure*}

In order to verify the tendency of users from the same region to be connected, we calculate the assortativity of the similarity networks. Assortativity measures the similarity of connections in the network with respect to a given attribute, and varies from $-$1 to +1~\cite{newman2002assortative}. In an \textit{assortative network} (with positive assortativity), vertices with similar values of the given attribute (e.g., same country) tend to connect with (be similar to) each other, whereas in a \textit{disassortative network} (with negative assortativity), the opposite happens. The assortativity analysis for the networks $G_{s}^1$ and $G_{s}^2$ formed from various values of $s$ are shown in Figures~\ref{fig:participatorySensorNet}b and \ref{fig:participatorySensorNet}c, respectively. Note that the assortativity for the network $G_{s}^1$ with respect to the geographical attributes (region Western/Eastern, continent, and country) decreases with the similarity threshold. This happens because most of the edges in 
the networks, formed from similarity threshold $s \geq$ 90, connect users who have preference vectors with a few positive features (as defined in Section~\ref{secMappingPreferences}). This also helps to explain why, in both figures, the degree assortativity increases with the similarity threshold: considering only very particular tastes, the network tends to be composed mostly of cliques, making the degree assortativity very close to 1.

On the other hand, if we vary the value of $s$ in the network $G_{s}^2$, the assortativity for geographical attributes remains roughly the same. It is possible to explain this behavior by looking at the size of the preference vector $F$ for the network $G_{s}^1$, which is much smaller compared to that for the network $G_{s}^2$ (101 against 435). Since the preferences are distributed over almost all the categories, a larger preference vector implies a lower probability of having preferences in common between two users, and, consequently, fewer edges in a similarity network, even for lower values of $s$. Note also that, in both Figures~\ref{fig:participatorySensorNet}b and \ref{fig:participatorySensorNet}c, all similarity networks we take into consideration are assortative. However, the assortativity values of the geographical attributes for $G_{s}^1$ are most of the time higher compared to those obtained for $G_{s}^2$. When considering all preferences/features we also increase the number of features that do 
not discriminate cultural differences sufficiently well (e.g., venues like homes, hotels, student centers, and shoe stores), since they are essentially present in all the cities and countries in the world. This suggests that, in this case, a similarity network considering only food and drink preferences might provide better insights in the study of cultural differences.

\begin{figure*}[t!]
\centering
\subfigure[Drink]
{\includegraphics[width=0.345\textwidth]{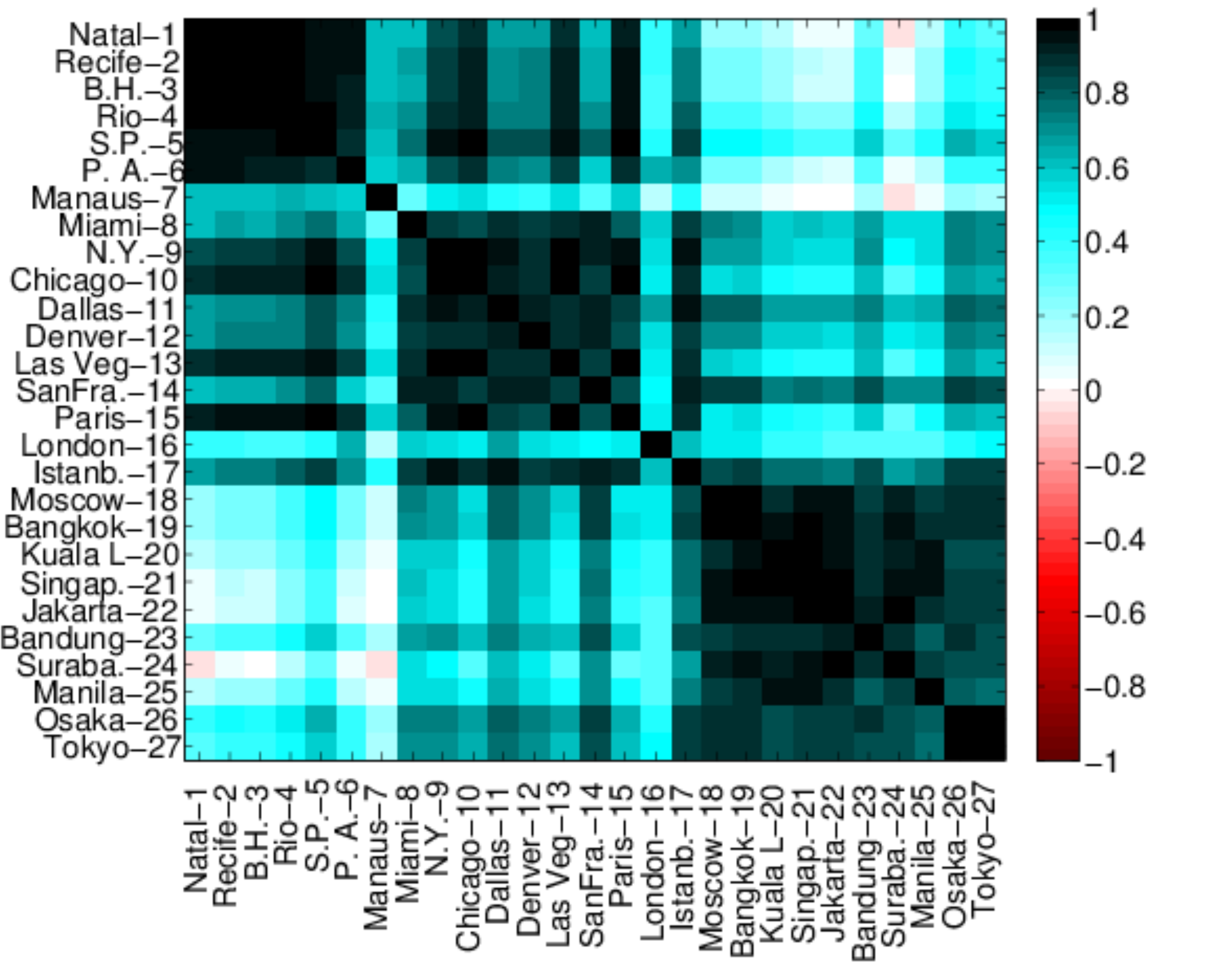}}
\subfigure[Fast Food]
{\includegraphics[width=0.31\textwidth]{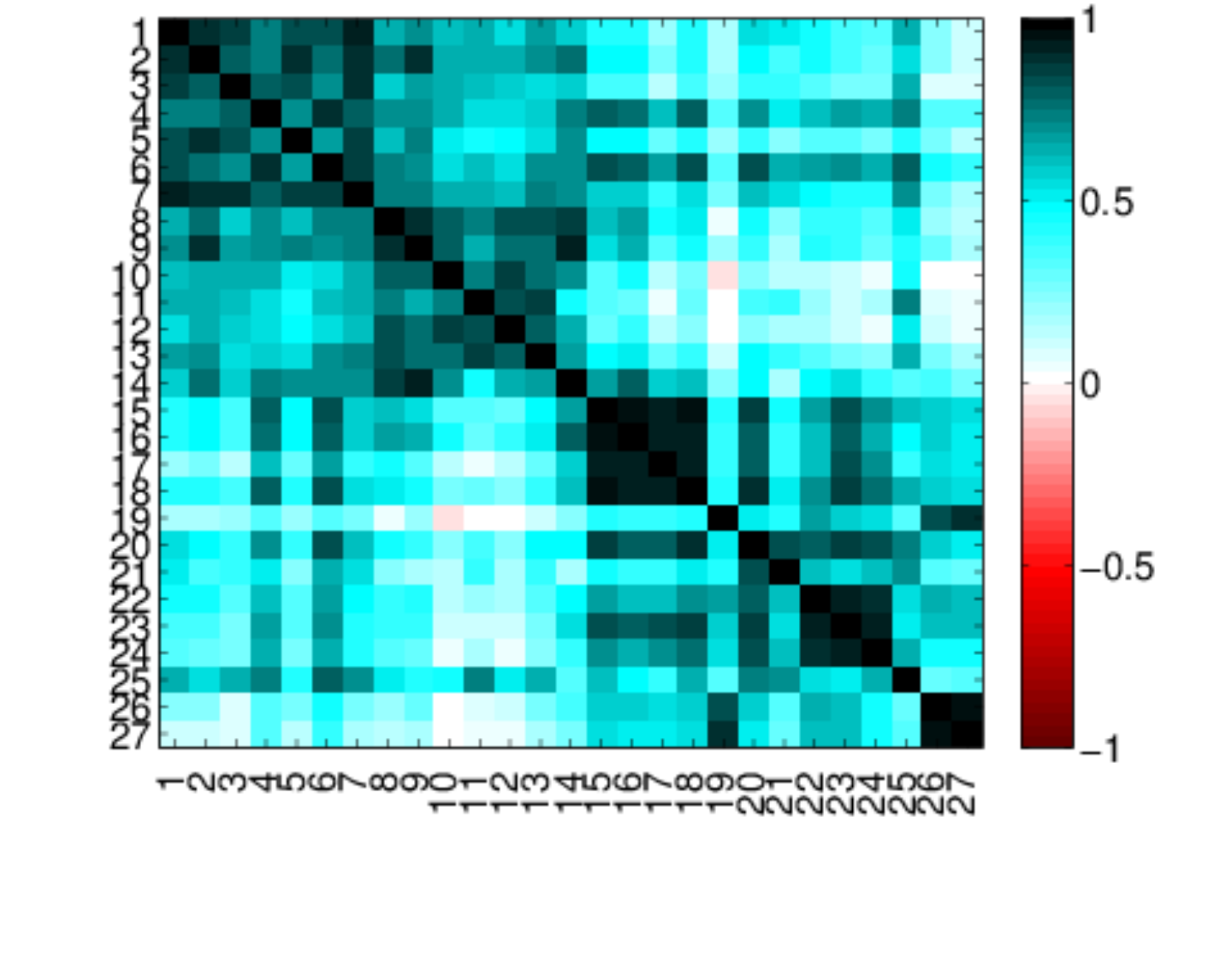}}
\subfigure[Slow Food]
{\includegraphics[width=0.31\textwidth]{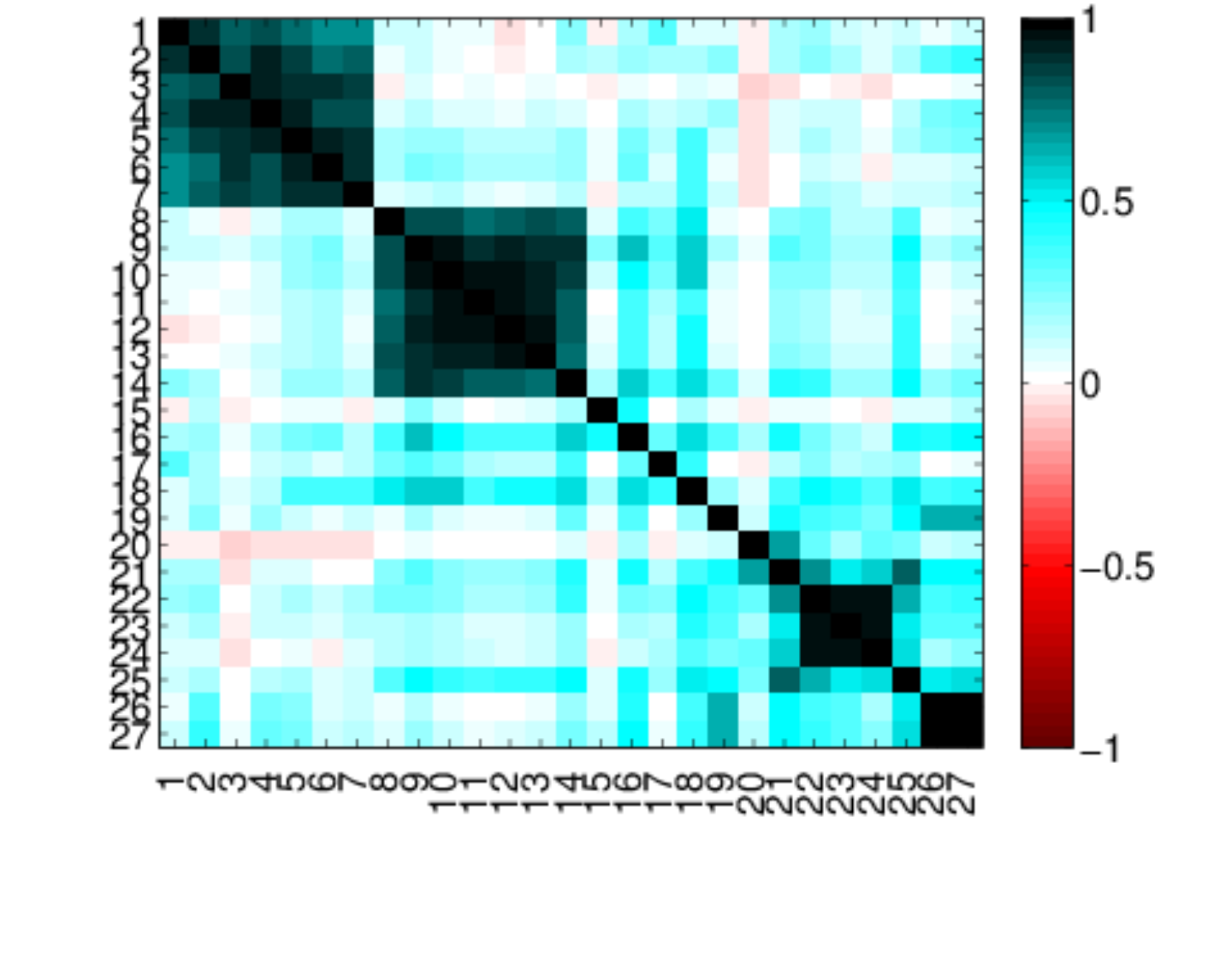}}\vspace{-4mm}
\caption{Correlation of preferences between cities.}\label{fig:correlaCities}
\end{figure*}

\section{Extraction of Cultural Signatures}\label{sec:featuresExtract}

Given the results discussed in Section~\ref{secIndividualPrefs}, we hypothesize that it is possible to define cultural signatures of different areas around the planet. In this section, we show how to extract features from Foursquare data that are able to describe regions from their cultural elements. In particular, we investigate two properties of food and drink preferences: their geographical (Section~\ref{secSpatialCorre}) and temporal (Section~\ref{secTemporalAna}) aspects.

\subsection{Spatial Correlations}\label{secSpatialCorre}

Here our goal is to define a set of features that are able to characterize the cultural preferences of a given geographical area in the planet, such as a country, a city or a neighborhood. Thus, for a given delimited area $a$ (e.g., the city of Chicago), we sum up the values of the features in the preference vectors of the users who checked in at venues of that area. In other words, we count the number of \checkins $C^a = c_1^a, c_2^a, \ldots, c_{101}^a$ performed in venues of each of the 101 subcategories $s_1, s_2, \ldots, s_{101}$ of the Fast Food, Slow Food and Drink classes (Section~\ref{sec:generalView}) that are located within the perimeter of area $a$. Next, we represent each area $a$ by a vector of $101$ features $F^a = f_1^a, f_2^a, \ldots, f_{101}^a$, where each feature $f_i^a$ is equal to $c_i^a/\max(C^a)$. That is, we normalize the number of \checkins at each subcategory by the maximum number of \checkins performed in a single subcategory in area $a$ ($\max(C^a)$). Thus, each area $a$ is 
represented by a feature vector $F^a$ containing values from 0 to 1, indicating the preferences of people who visited that area, i.e., the profile of preferences for that area. From now on, we use $F_{drink}^a$, $F_{sfood}^a$ and $F_{ffood}^a$ to refer, respectively, to the subset of features that correspond to subcategories belonging to the Drink, Slow Food and Fast Food classes in area $a$.

In order to verify if two areas $a$ and $b$ are culturally similar, we compute the Pearson's correlation coefficient between the two feature vectors $F^a$ and $F^b$ of those areas. We compute the correlation considering all features ($F^a$ and $F^b$) as well as a subset of them (e.g., $F_{drink}^a$ and $F_{drink}^b$). In particular, Figure \ref{fig:correlationCountries} shows the correlations between areas corresponding to 27 different popular countries for the Drink (\ref{fig:correlationCountries}a), Fast Food (\ref{fig:correlationCountries}b), and Slow Food~(\ref{fig:correlationCountries}c) classes; the darker the color, the stronger the correlation (blue for positive correlations, red for negative correlations). The same correlations computed for city level areas (16 cities around the world) are shown in Figure~\ref{fig:correlaCities}.

Analyzing the results for the Drink class (Figure \ref{fig:correlationCountries}a), we find countries with very strong correlations, such as Argentina and Chile, as well as countries with low correlation, such as Brazil and Indonesia. Moreover, although regions close geographically tend to have stronger correlations, this is not always the case. For example, the correlation between Brazil and France is stronger than the correlation between England and France, which are geographically closer. Similarly, Figure~\ref{fig:correlaCities}a\footnote{The ratio of \checkins per inhabitant is similar among all the cities taken into consideration. For example, comparing Manaus (one of the cities with fewer \checkins) with Sao Paulo (largest number of \checkins in Brazil) we find the following ratios: $0.35\times10^{-3}$ and $0.37\times10^{-3}$ (Drink class); $0.73\times10^{-3}$ and $0.75\times10^{-3}$ (Fast Food class); and $0.54\times10^{-3}$ and $0.71\times10^{-3}$ (Slow Food class).} shows that cities in the same 
country tend to have very correlated drinking habits in most cases, but there are exceptions: Manaus (Brazil), for instance, has weak correlation with other cities in Brazil. This might be due to this city being located in the North region of Brazil, which is known for having a strong cultural diversity compared to other parts of the country.

Turning our attention to food practices, we observe in Figures~\ref{fig:correlationCountries}b and \ref{fig:correlaCities}b the global penetration of fast food venues, at both country and city levels, explained by the diffusion of fast food places worldwide~\cite{watson2006golden}. This is not observed in the same intensity for the Slow Food class (Figures~\ref{fig:correlationCountries}c and \ref{fig:correlaCities}c). The Slow Food class presents the highest distinction, or smaller correlation, across most of the countries and cities. This is expected, since Slow Food venues usually are representative of the local cuisine. Note, for instance, that cities from Brazil and USA have highly correlated drinking and fast food habits, but almost no correlation in slow food habits.

\begin{figure}[t!]
\centering
\subfigure[NY]{\includegraphics[width=37mm,height=40mm]{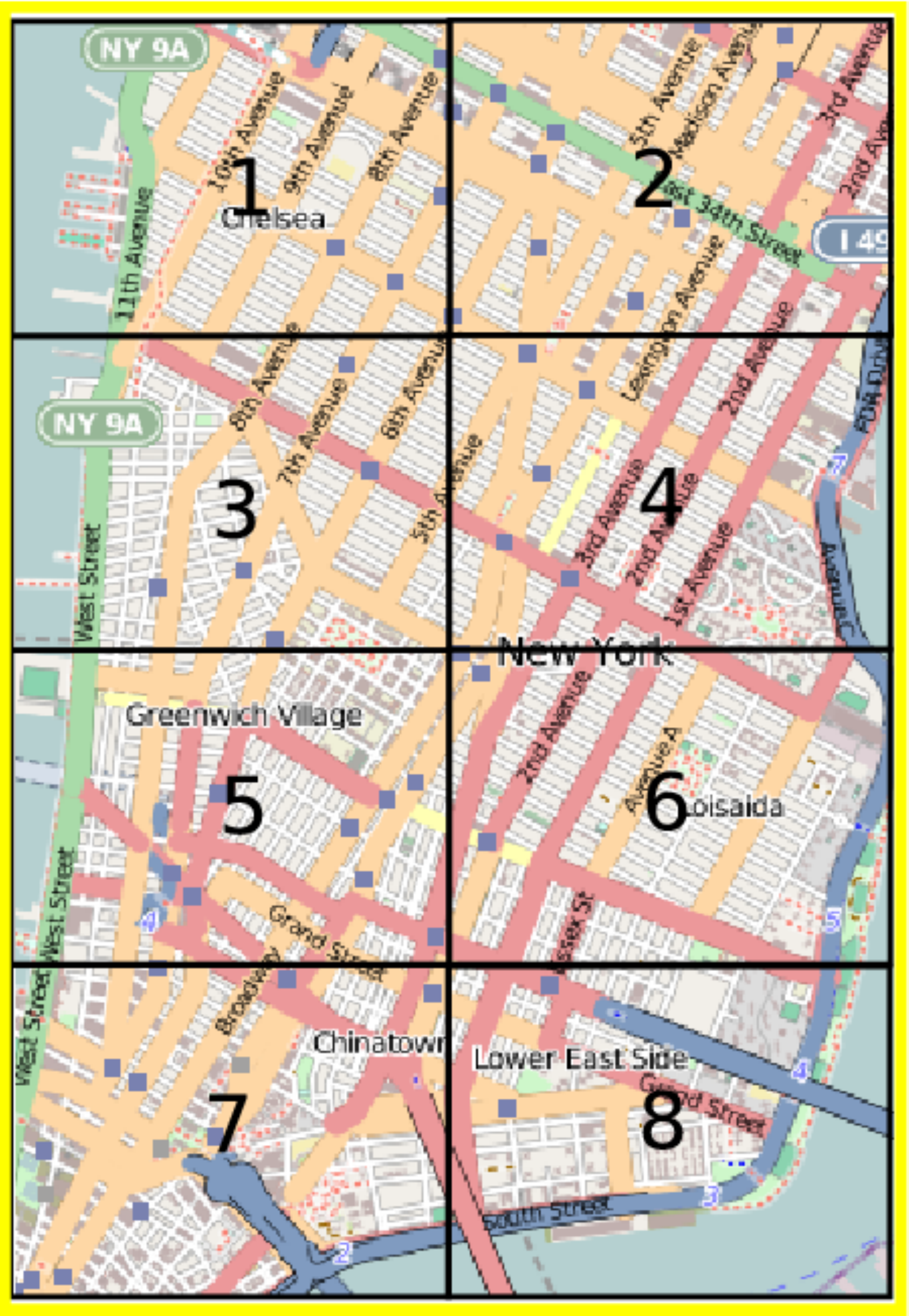}}
\subfigure[Tokyo]{\includegraphics[width=37mm,height=40mm]{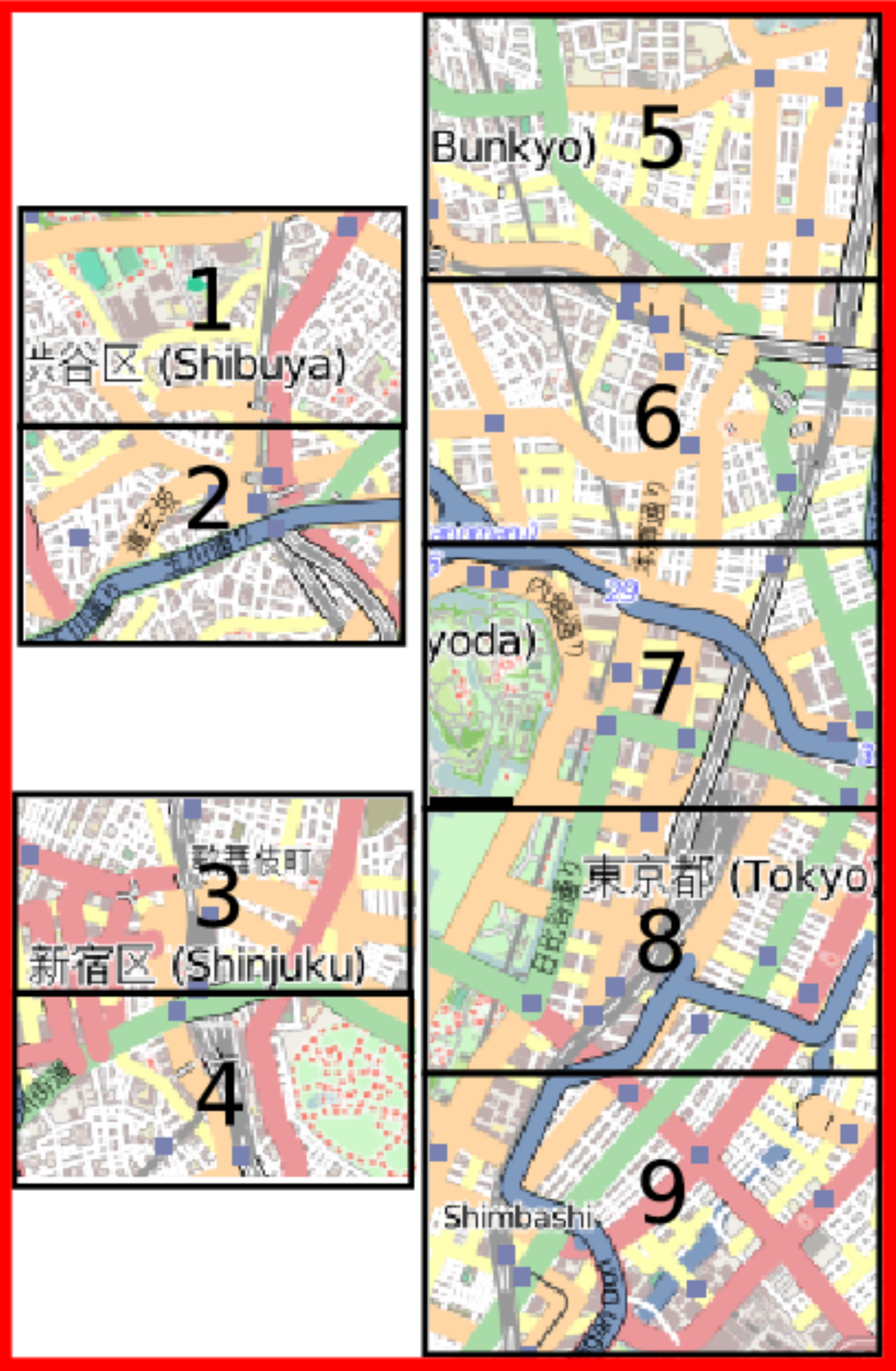}}
\subfigure[London]{\includegraphics[width=49mm]{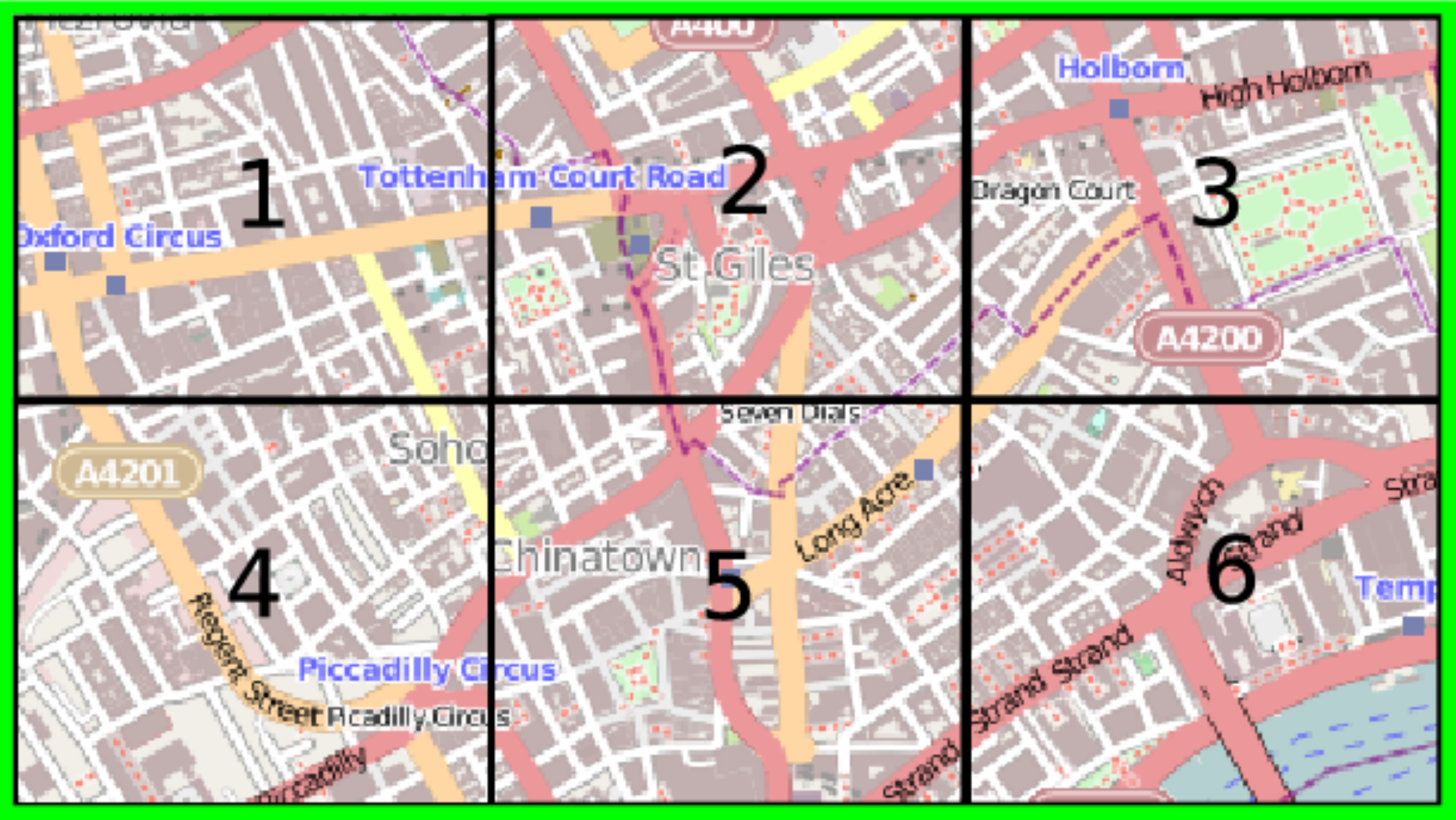}}\vspace{-4mm}
\caption{Areas of cities taken into consideration: London/England; New York/USA; and Tokyo/Japan.}\label{fig:citiesAreas}
\end{figure}

Finally, we turn our attention to the cultural habits within city boundaries. It is known that, in many cities, there is a strong cultural diversity across different neighborhoods~\cite{cranshaw:livehoods}, reflecting distinct activities typically performed in these areas. To analyze these local cultures, we focus on three populous cities, namely London, New York, and Tokyo. We divide each city's geographical area using a grid structure. Next, we select the most popular cells in the grid of each city and label them with a number, as shown in Figure~\ref{fig:citiesAreas}. We then compute the correlation between the selected cells. Note that we here assume a grid with regular (rectangular) cells to show the potential of the proposed analysis. However, our approach can be applied to any other segmentation of the city areas (e.g., by city districts).

\begin{figure*}[t!]
\centering\subfigure[Drink]
{\includegraphics[width=0.32\textwidth]{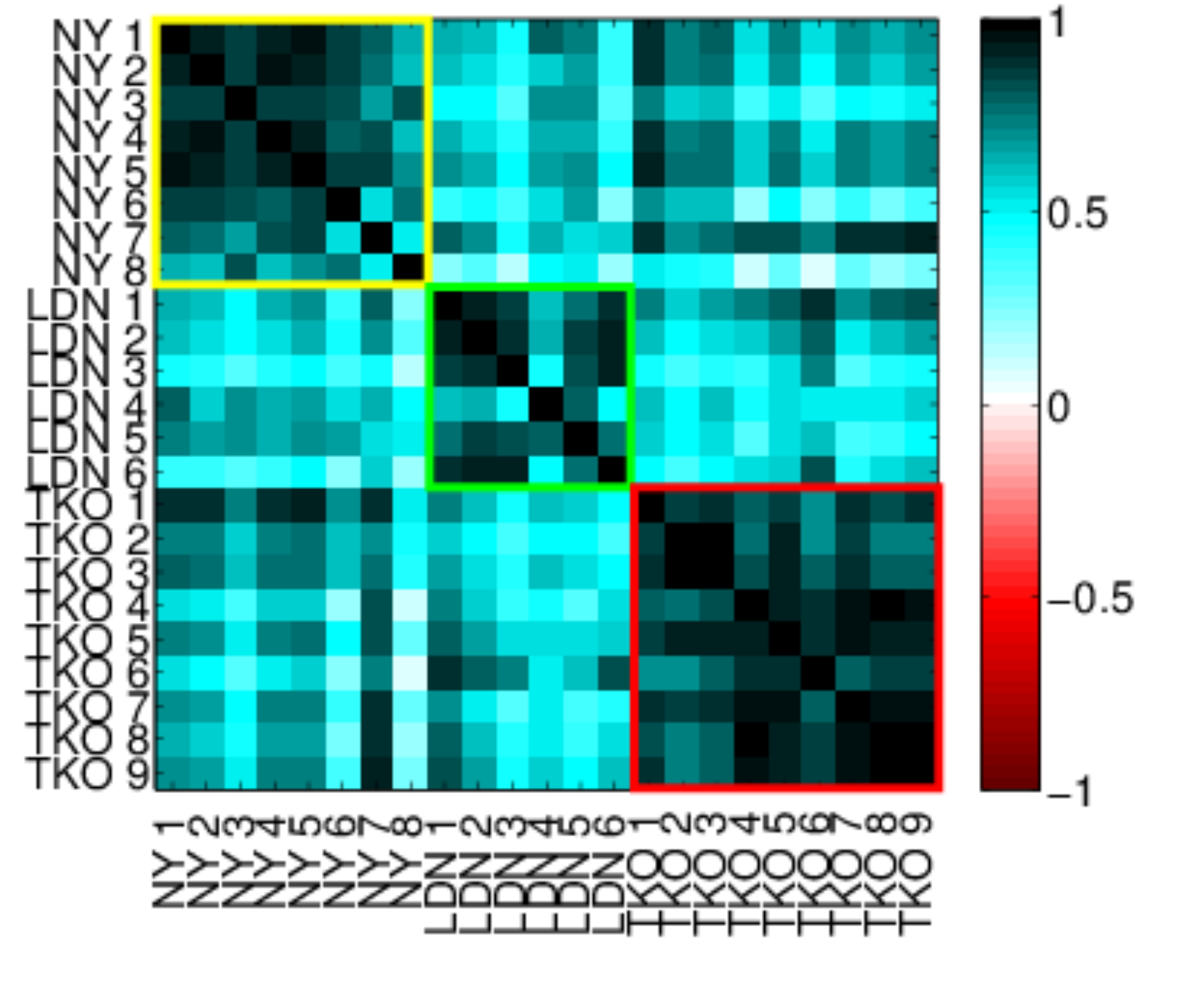} }
\centering\subfigure[Fast Food]
{\includegraphics[width=0.32\textwidth]{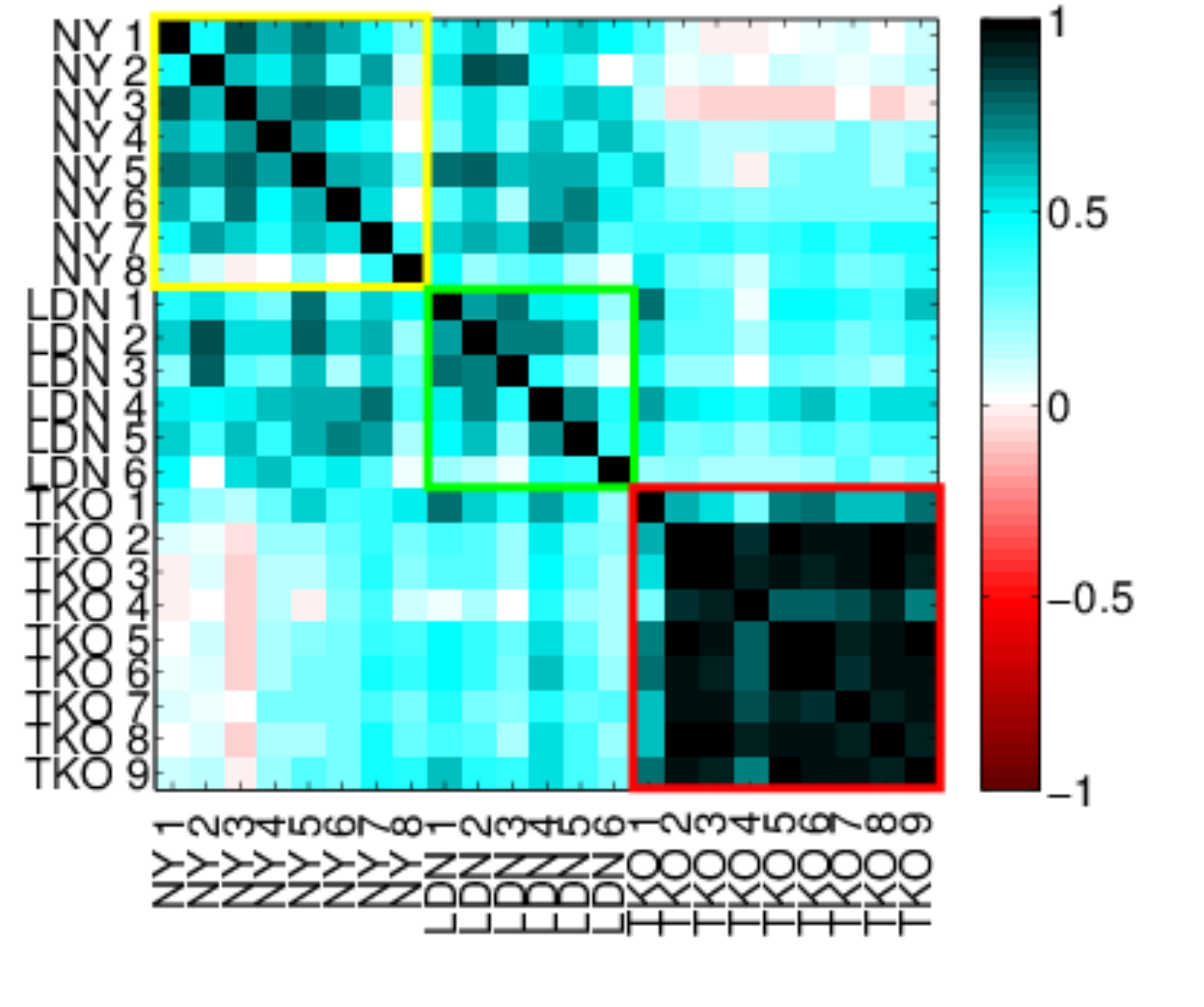} }
\centering\subfigure[Slow Food]
{\includegraphics[width=0.32\textwidth]{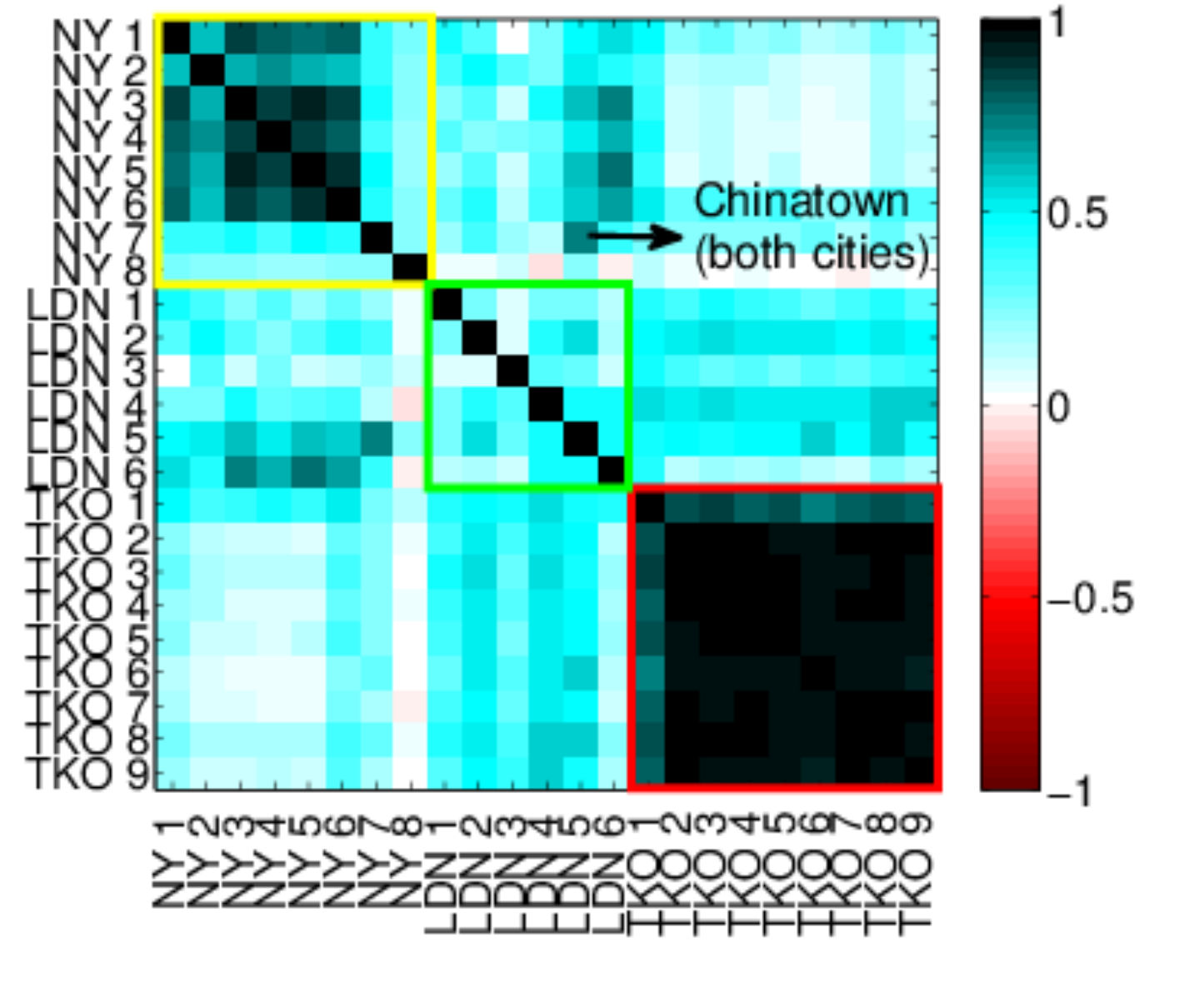} }\vspace{-4mm}
\caption{Correlation of preferences in regions of London, NYC and Tokyo.}\label{fig:correlationAreas}
\end{figure*}

Figure~\ref{fig:correlationAreas} shows the correlations for pairs of cells within the same city and from different cities. Note that, for the Drink class, different areas within the same city tend to have very strong correlations. There are also areas from different cities with strong correlations (e.g., areas NY-5 and TKO-1). For Fast Food places, the correlations between areas within the same city are much stronger for Tokyo, although the correlations between New York and London areas are fairly moderate. In contrast, there are areas with negative correlation, e.g., NY-3 with most of Tokyo areas.

Finally, for the Slow Food class, once again Tokyo areas are very strongly correlated among themselves. In comparison with the Fast Food class, there is a more clear distinction (weaker correlation) between London and New York areas as well as among distinct areas in London. This last observation is probably due to a specific characteristic of London, that has neighborhoods with a strong presence of a cuisine of a particular region of the globe. Observe also that two specific areas of New York, namely NY-7 and NY-8, are particularly not correlated with the others from this city. This is probably related to the location of Chinatown in those areas (mainly NY-7). Indeed, this particular area (NY-7) has a strong correlation with a particular area of London, LND-5, where Chinatown/London is located.

\subsection{Temporal Analysis}\label{secTemporalAna}

We now turn our attention to the temporal and circadian aspects of cultural habits. The time instants when \checkins are performed in food and drink places may also provide valuable insights into the cultural aspects of a particular region. For example, in a particular area, one may like to drink beer during the weekends but not during the weekdays.

\begin{figure}[t!]
\centering
\subfigure[Drink, WD]
{\includegraphics[width=0.15\textwidth]{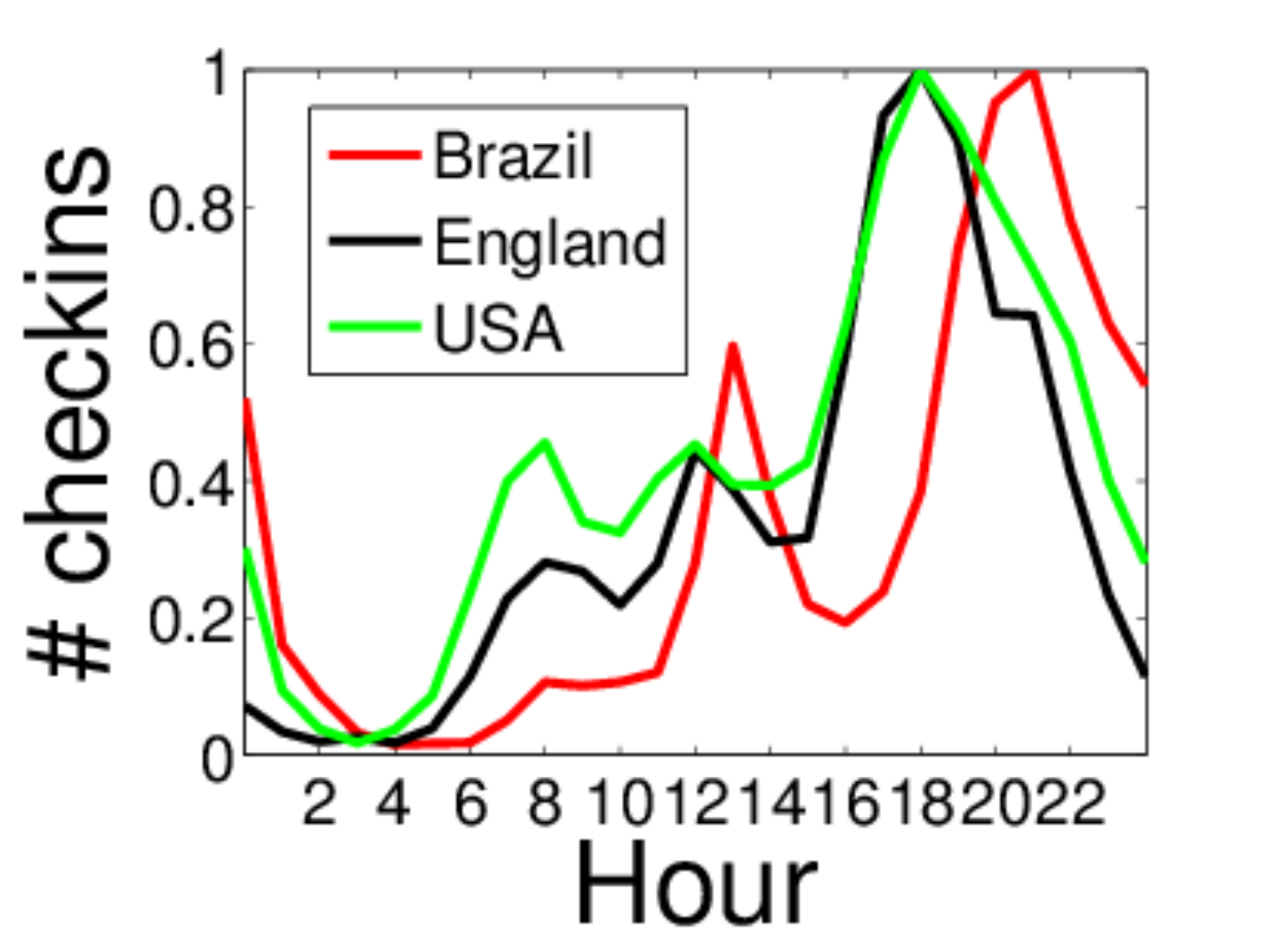}}
\subfigure[Fast Food, WD]
{\includegraphics[width=0.15\textwidth]{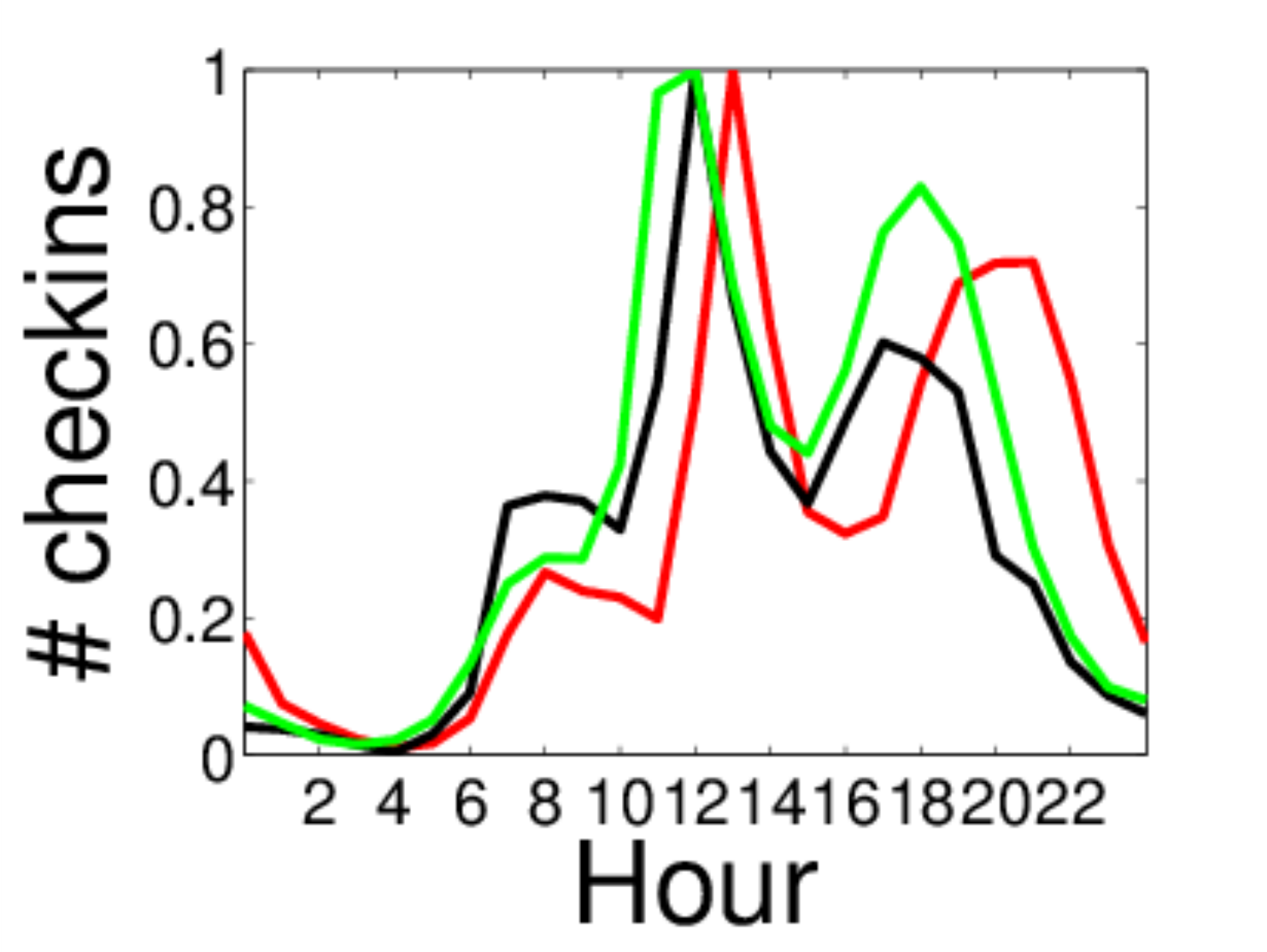}}
\subfigure[Slow Food, WD]
{\includegraphics[width=0.15\textwidth]{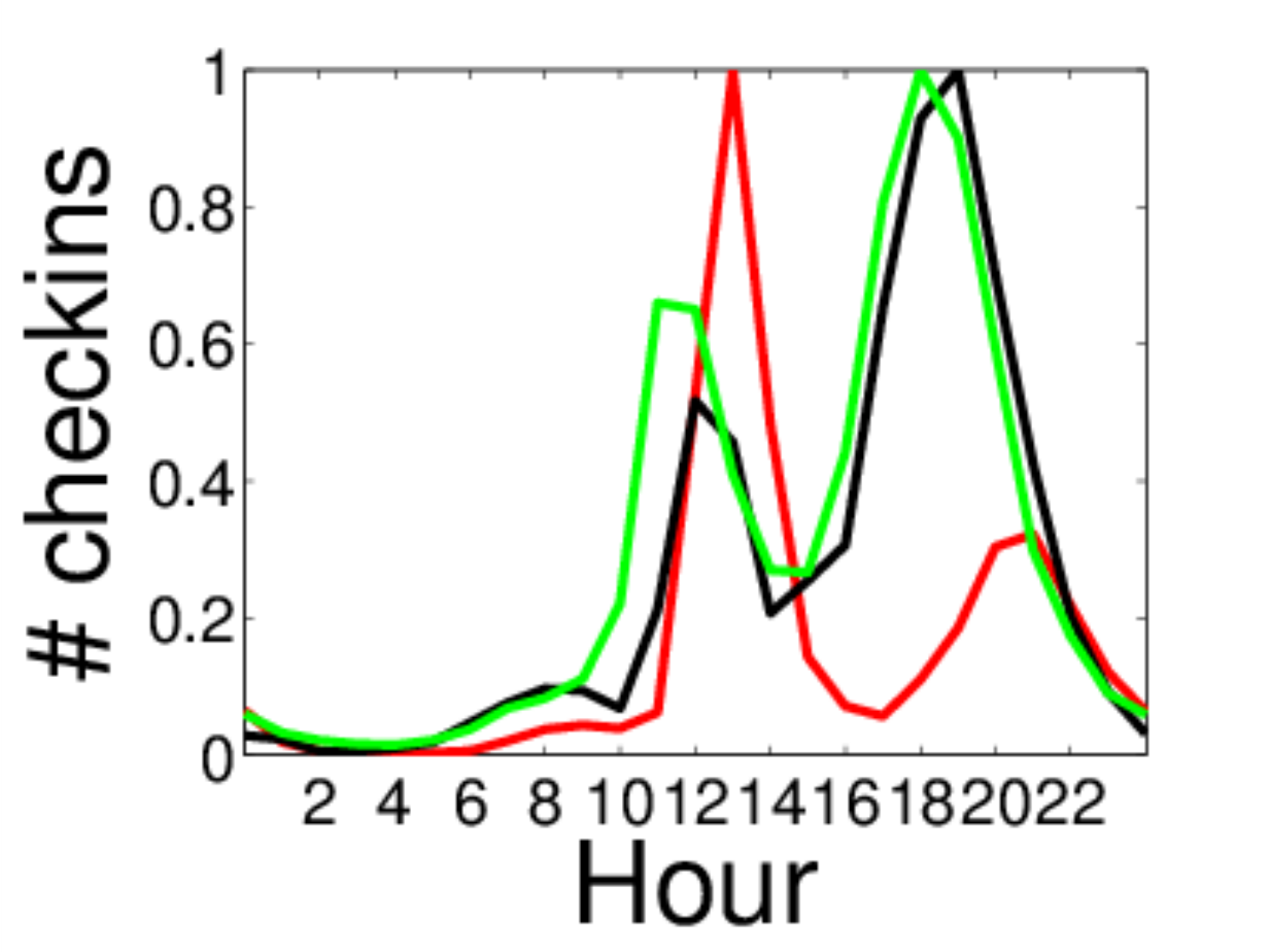}}
\subfigure[Drink, WE]
{\includegraphics[width=0.15\textwidth]{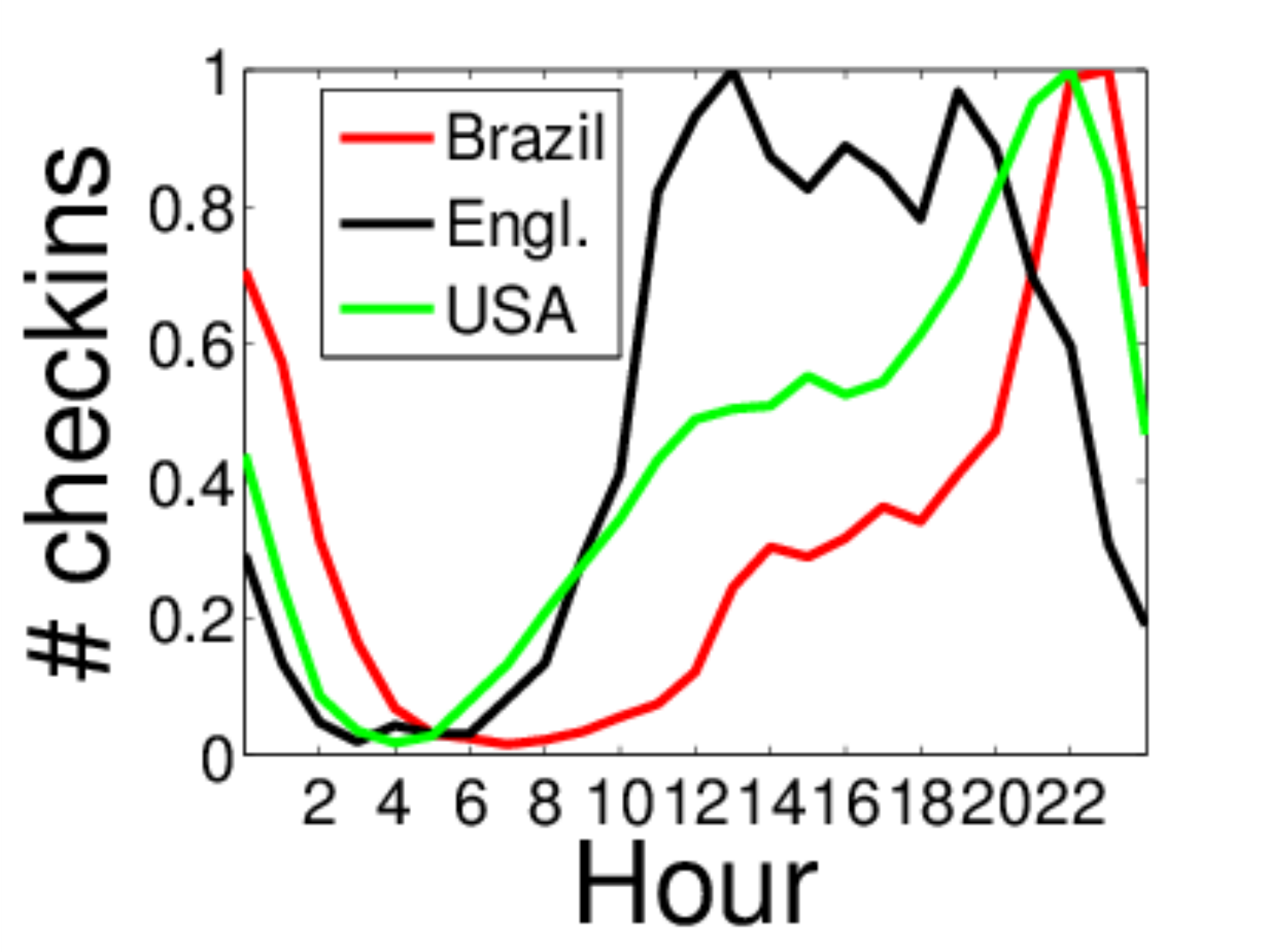}}
\subfigure[Fast Food, WE]
{\includegraphics[width=0.15\textwidth]{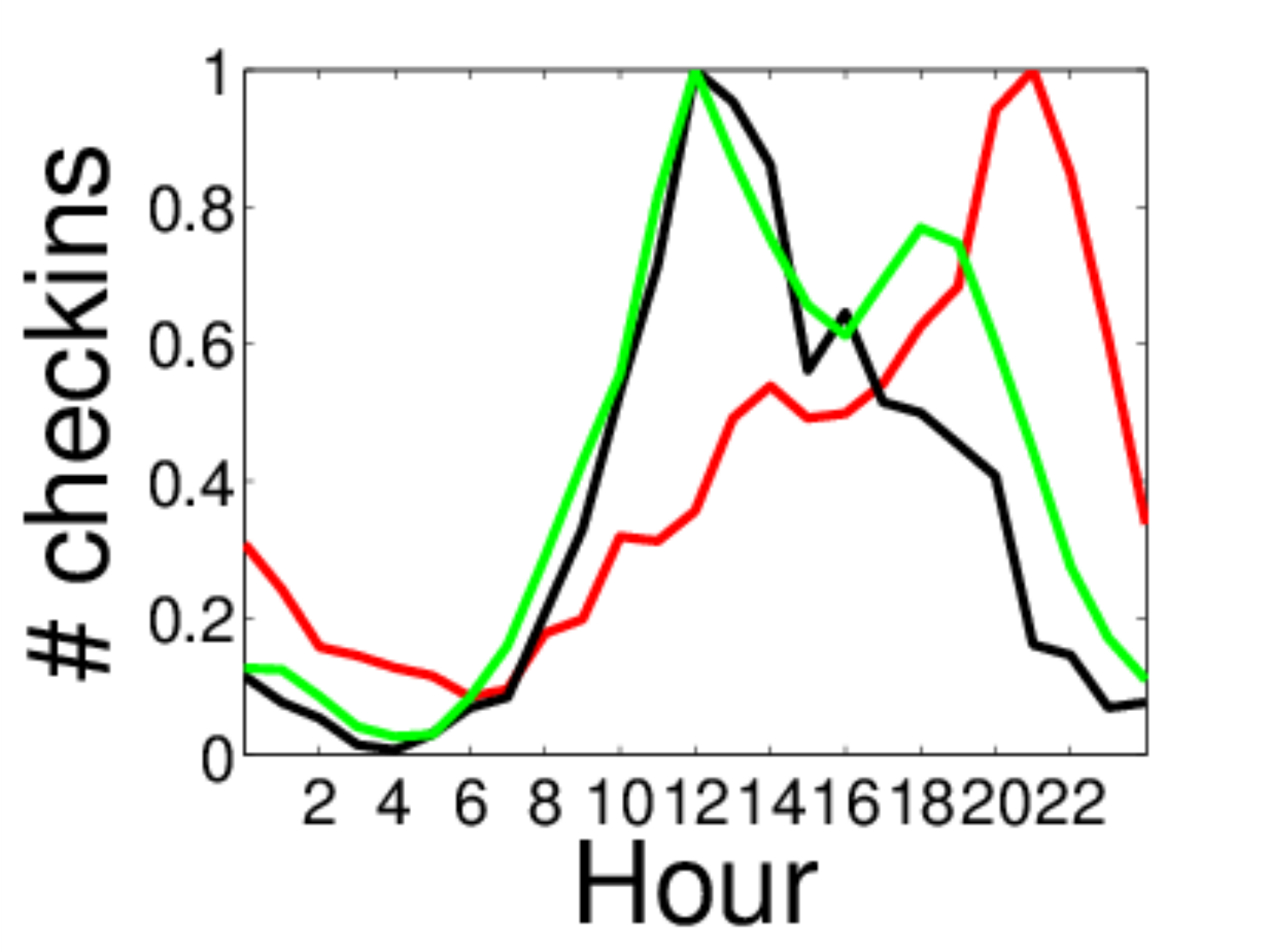}}
\subfigure[Slow Food, WE]
{\includegraphics[width=0.15\textwidth]{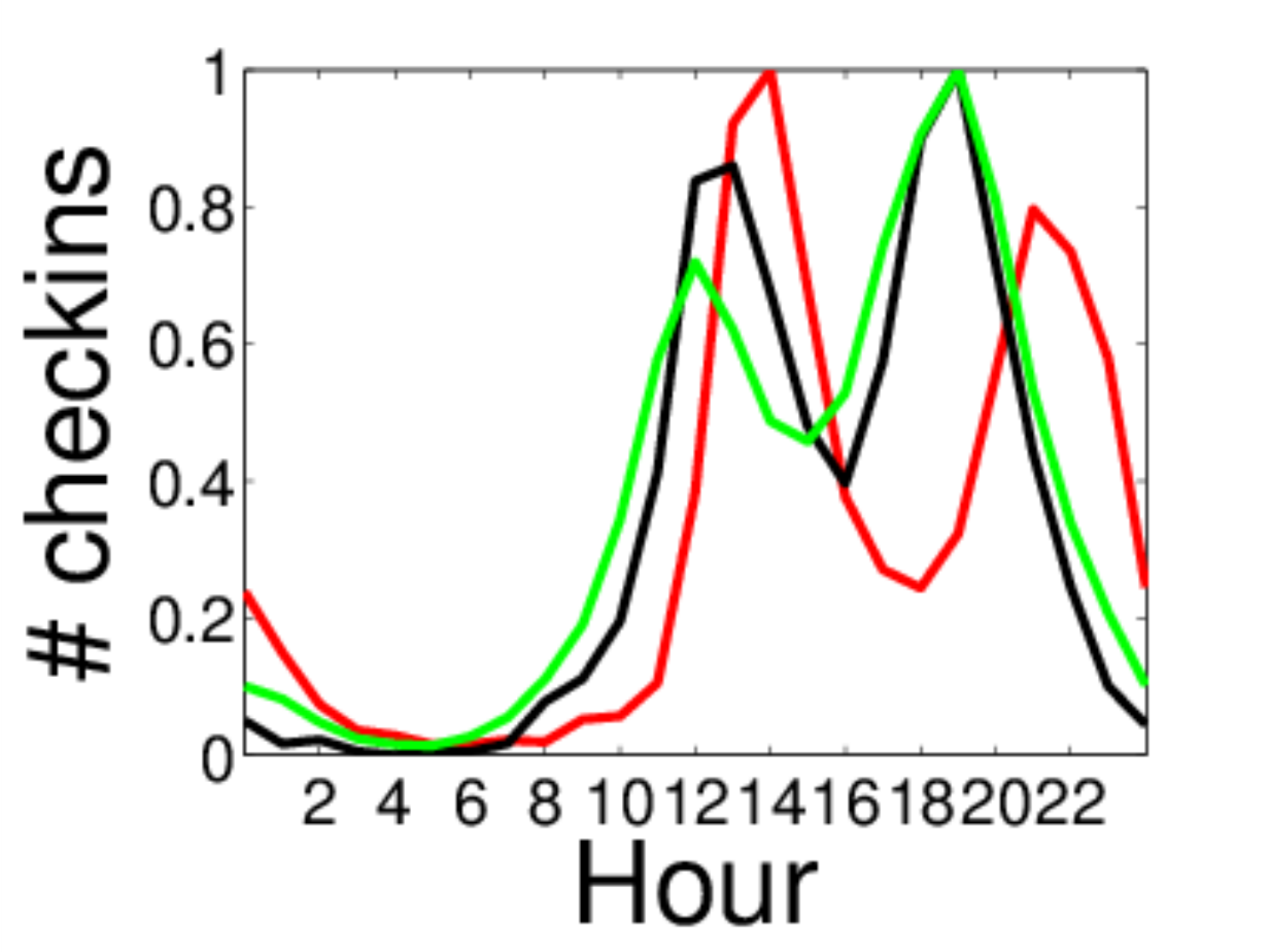}}\vspace{-4mm}
\caption{\# of \checkins throughout the hours of the day in different countries (WD = weekday; WE = weekend).}\label{fig:chkCountryWeekdayWeekend}
\end{figure}

\begin{figure}[t!]
\centering
\subfigure[Drink, WD]
{\includegraphics[width=0.15\textwidth]{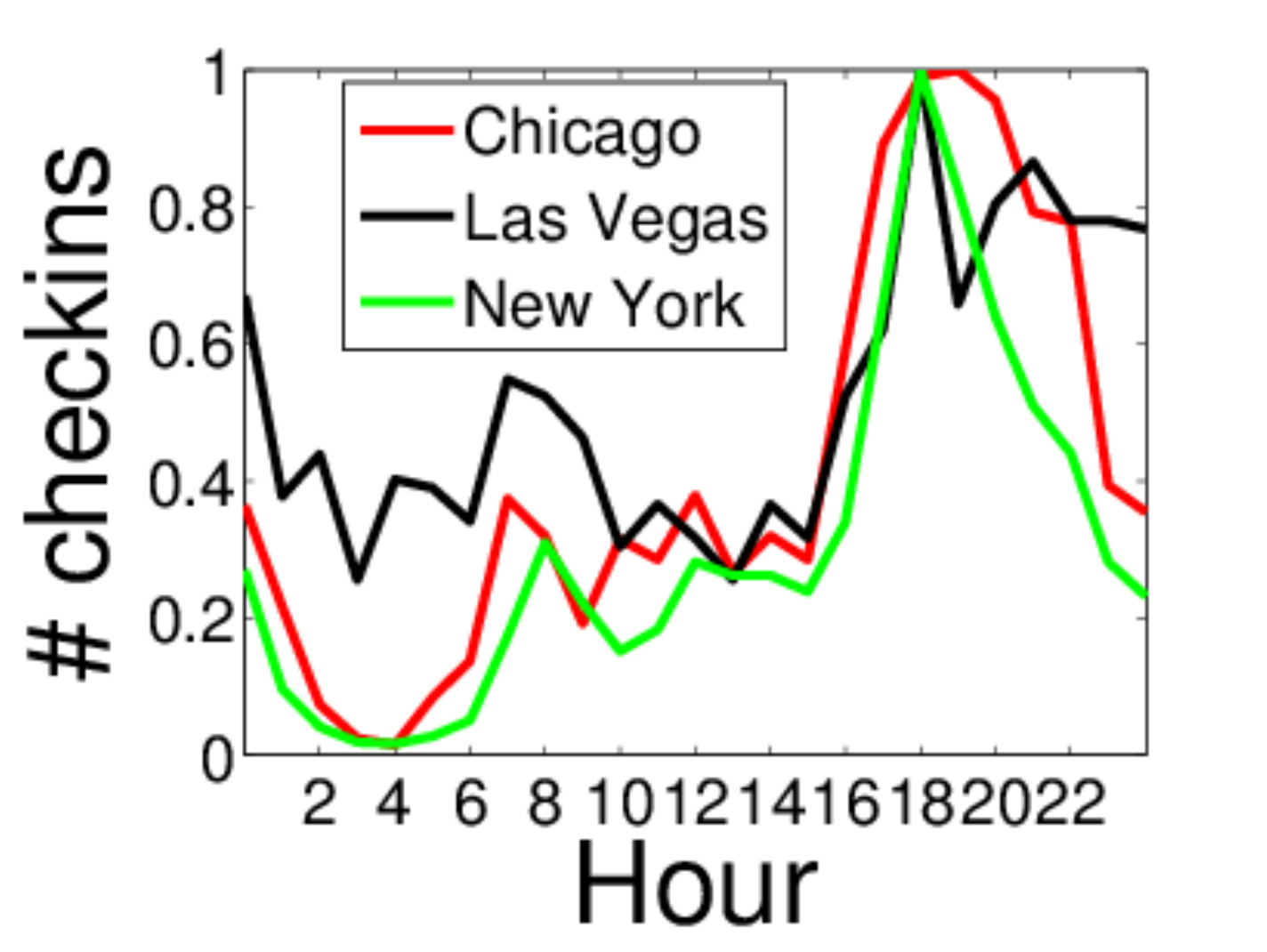}}
\subfigure[Fast Food, WD]
{\includegraphics[width=0.15\textwidth]{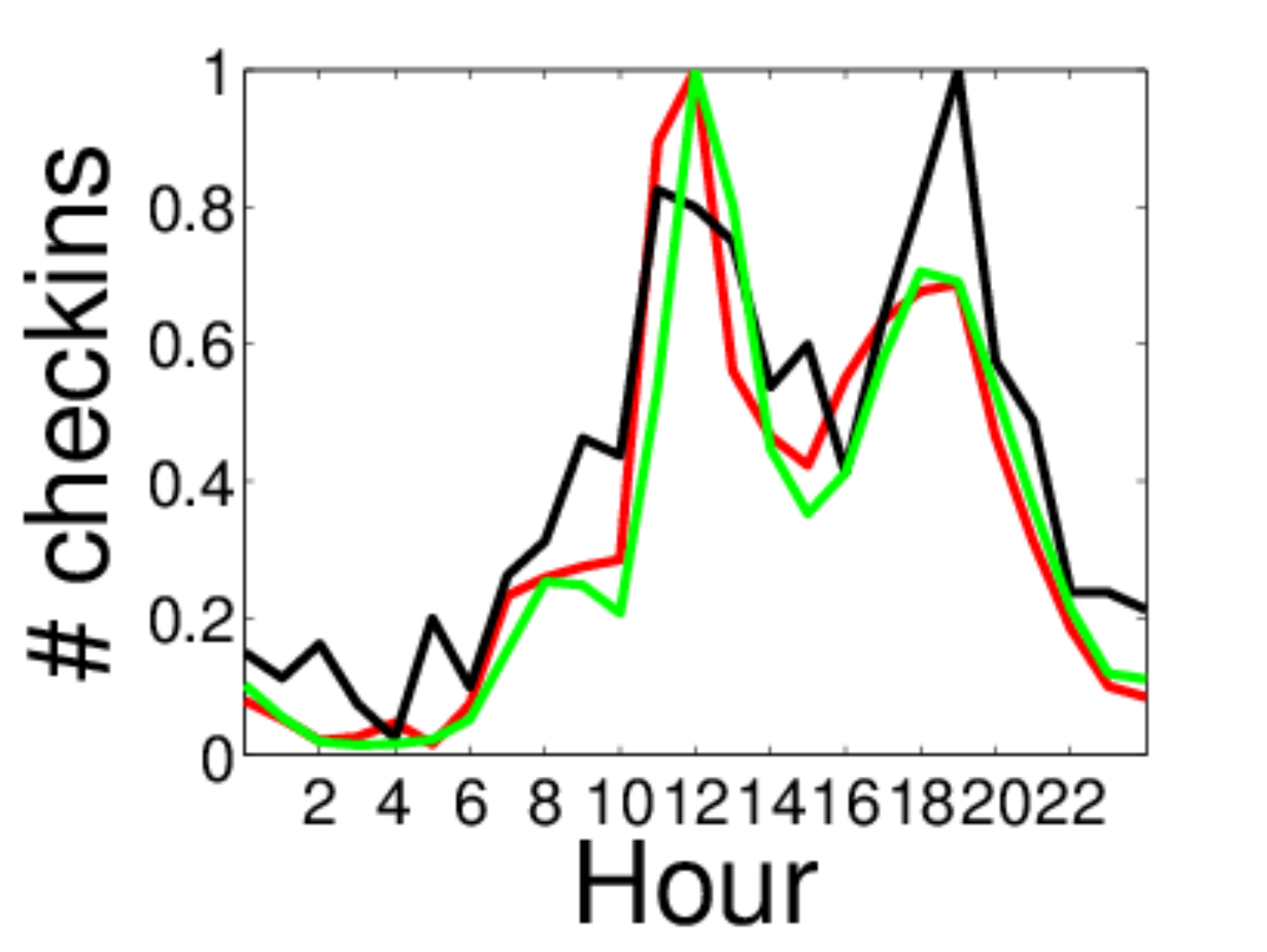}}
\subfigure[Slow Food, WD]
{\includegraphics[width=0.15\textwidth]{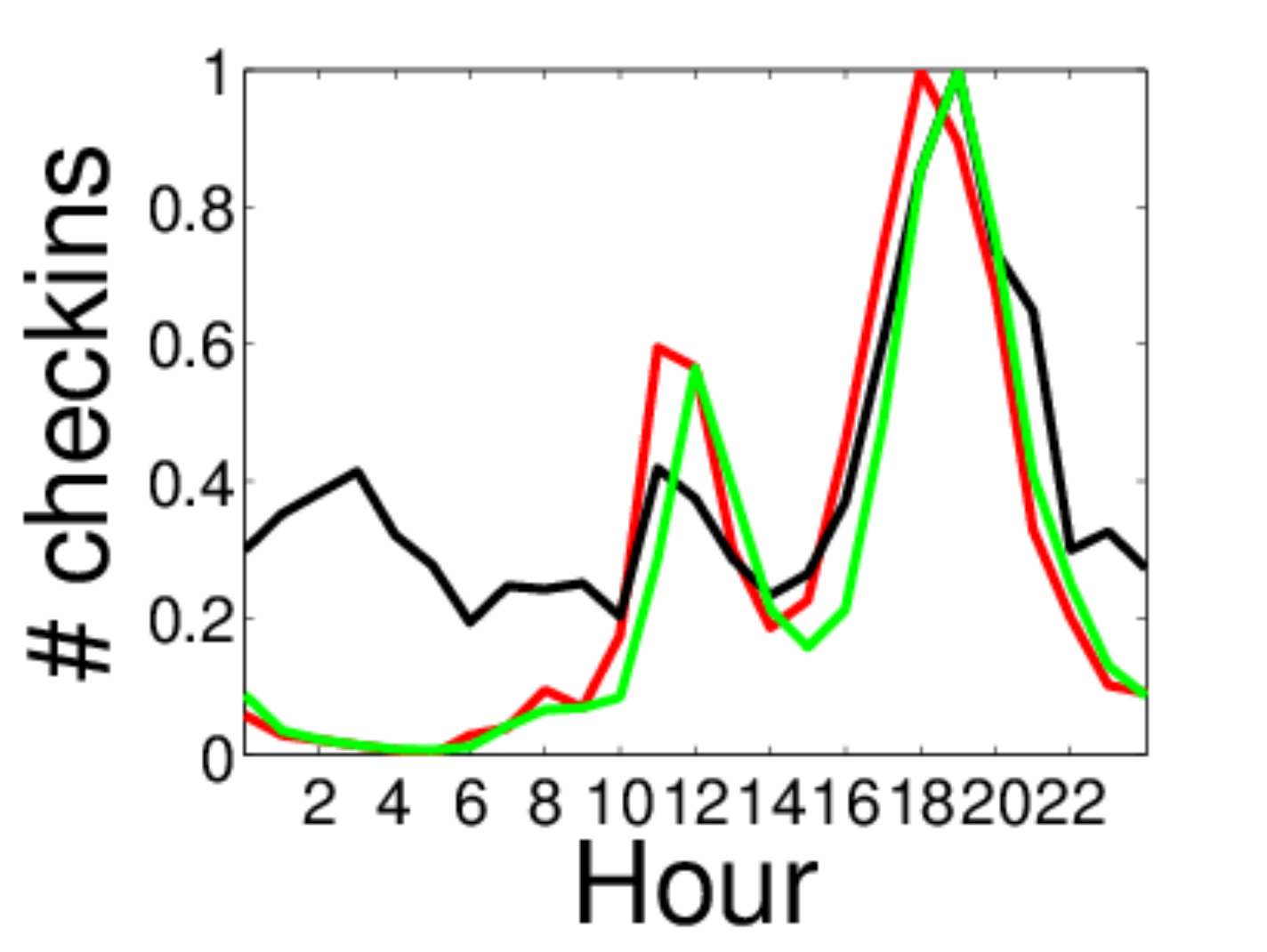}}
\subfigure[Drink, WE]
{\includegraphics[width=0.15\textwidth]{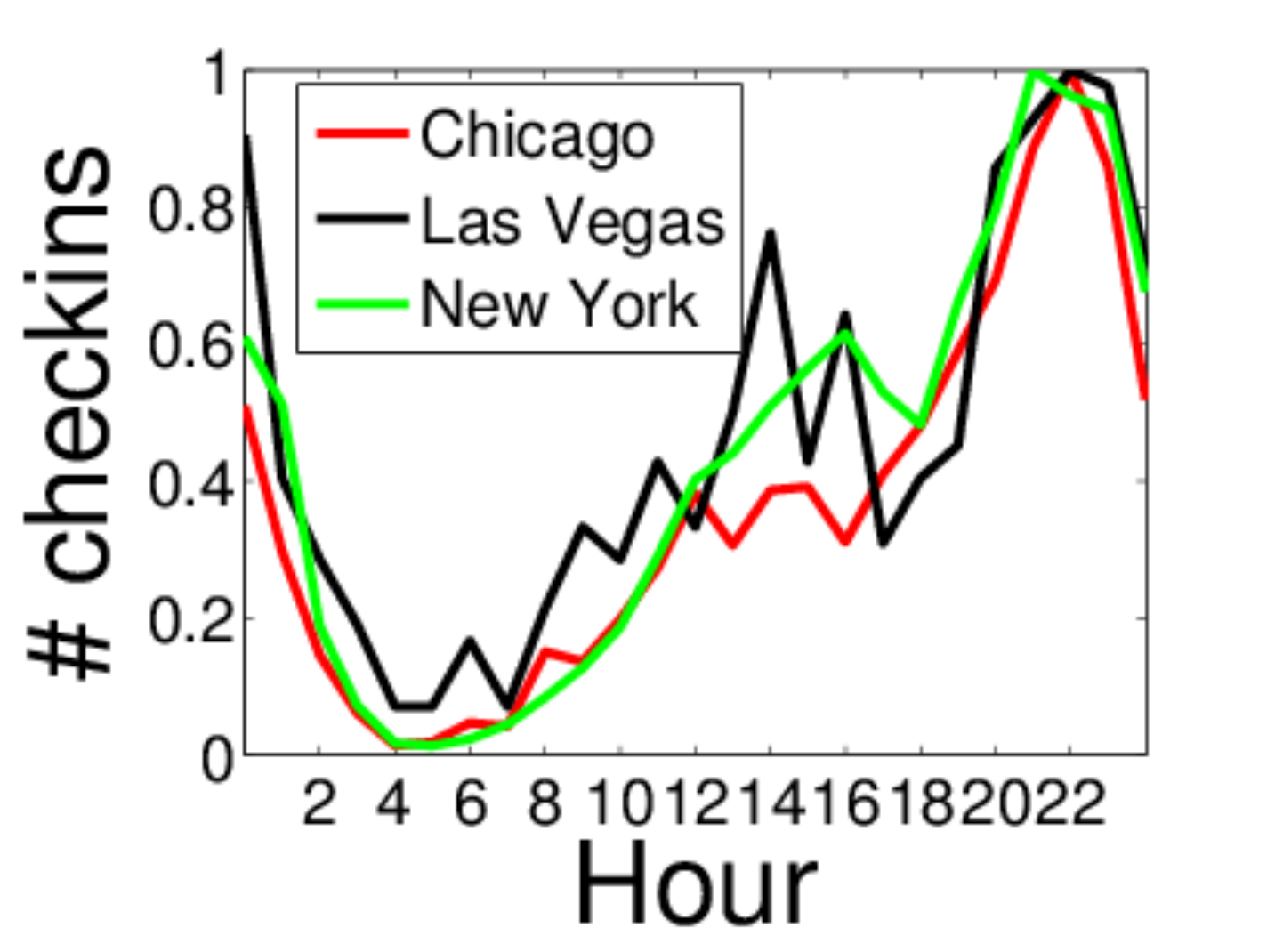}}
\subfigure[Fast Food, WE]
{\includegraphics[width=0.15\textwidth]{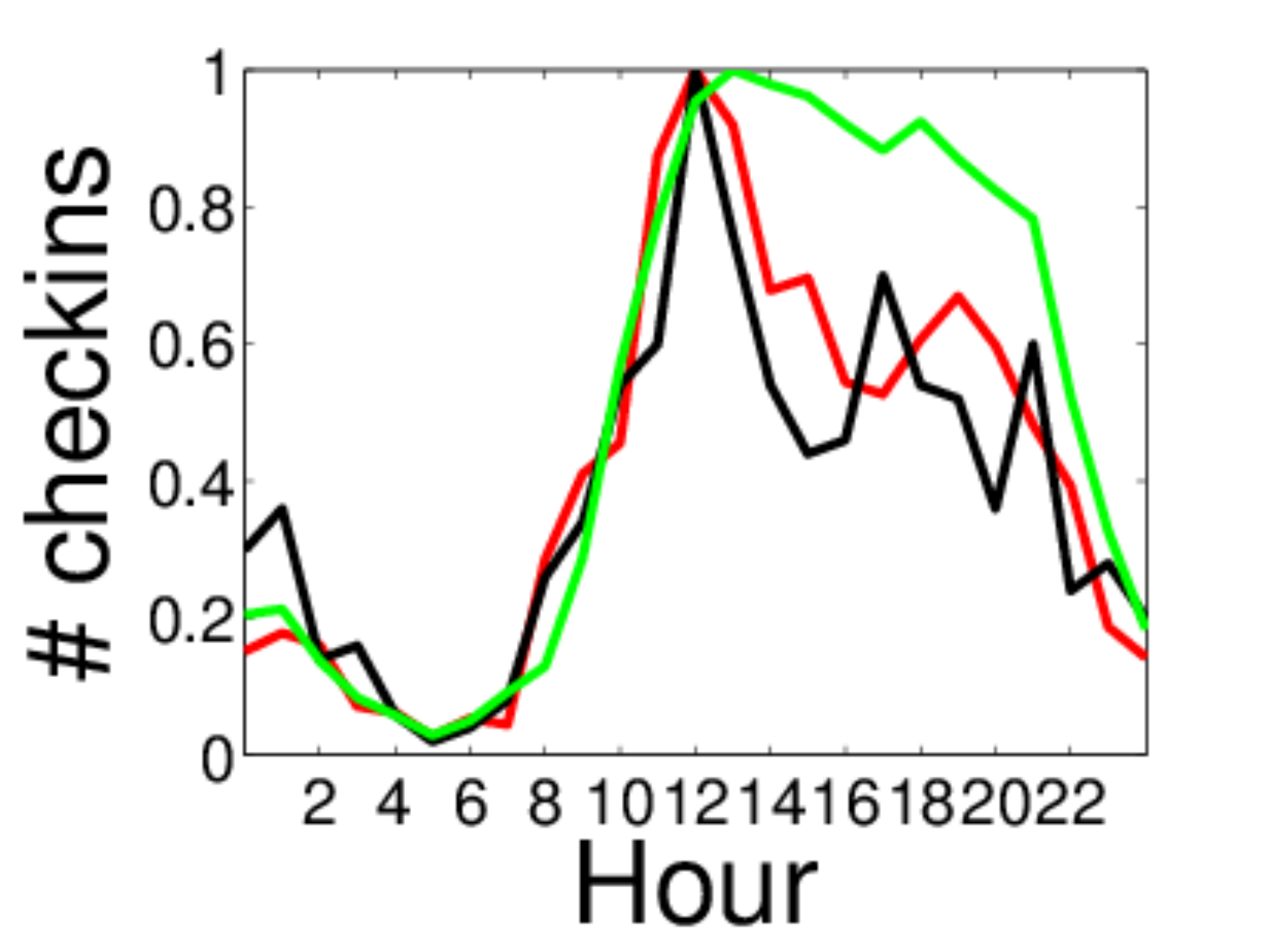}}
\subfigure[Slow Food, WE]
{\includegraphics[width=0.15\textwidth]{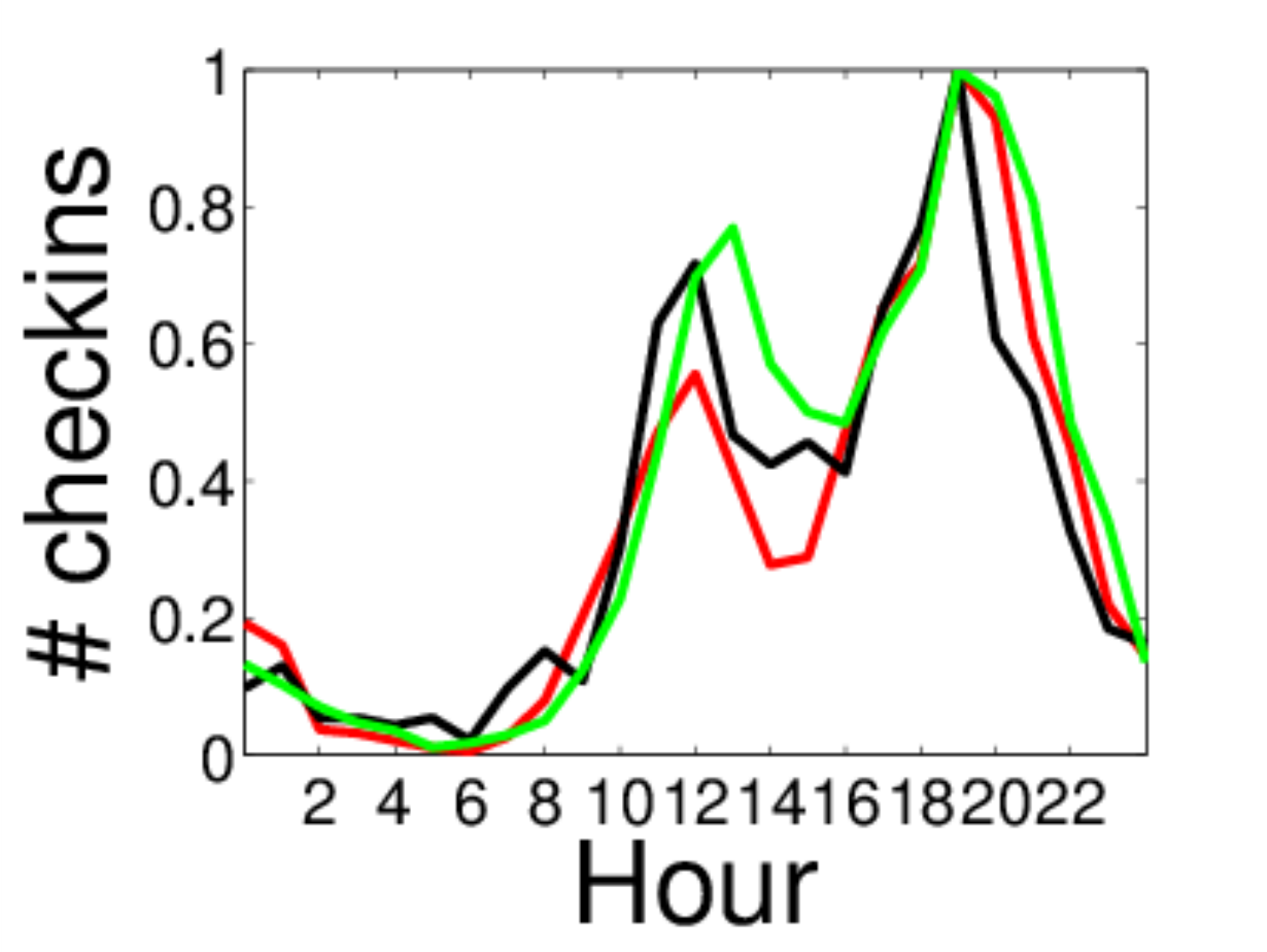}}\vspace{-4mm}
\caption{\# of \checkins throughout the hours of the day in different American cities (WD = weekday; WE = weekend).}\label{fig:chkCityWeekdayWeekend}
\end{figure}

To that end, we first count the number of \checkins per hour during the whole week covered by our dataset in venues of each class (Drink, Fast Food and Slow Food) for different regions. Next, we group days into weekdays and weekends, summing up the \checkins performed on the same hour of the day in each group and for each region. We then normalize this number by the maximum value found in any hour for the specific region, so that we can compare the patterns obtained in different regions. For illustration purposes, we show the results for three countries (Brazil, USA, and England) and for three American cities (Chicago, Las Vegas, and New York) in Figures \ref{fig:chkCountryWeekdayWeekend} and \ref{fig:chkCityWeekdayWeekend}, respectively. Results for each class are shown separately for weekdays and weekends.

Focusing first on weekday patterns, Figure \ref{fig:chkCountryWeekdayWeekend} shows that American and English people have similar peaks of activities, despite differences in their preferences for different categories of places, as previously shown (Figure~\ref{fig:correlationCountries}). In contrast, Brazilians tend to have significantly different temporal patterns, particularly in terms of activities in Slow Food places (Figure \ref{fig:chkCountryWeekdayWeekend}c): whereas Americans and English people tend to have their main meal at dinner time, Brazilians have it at lunch time. Observe also that Brazilians have their meals later, compared to Americans and English people.

Concerning the times when people go to drink venues, it is possible to note similarities among most of the cities from the same country, but also some different patterns. For example, most of the analyzed cities from USA exhibit a weekday pattern similar to New York and Chicago, shown in Figure~\ref{fig:chkCityWeekdayWeekend}a, with three distinct peaks around breakfast, lunch and happy hour (around 6pm). This behavior is consistent with the general pattern observed for the country, shown in Figure~\ref{fig:chkCountryWeekdayWeekend}a. However, Las Vegas is one exception, since there is an intense activity during the dawn, besides many other peaks of activities that do not occur in other cities.

Turning our attention to eating habits on weekdays, Figure \ref{fig:chkCityWeekdayWeekend} shows that most cities in the USA present activity patterns very similar to the general pattern identified for the country, both in terms of Slow and Fast Food places. However, as observed for drinking patterns, there are exceptions, such as Las Vegas, which exhibits distinct trends that reflect inherent idiosyncrasies of this city. We also note relevant similarities and differences in eating habits of people from cities in different countries. For example, comparing Figures~\ref{fig:chkCityWeekdayWeekend}b and~\ref{fig:chkCityWeekdayWeekend}c with similar graphs produced for different Brazilian cities, we find that while all curves for the Fast Food class are very similar, the curves for Slow Food places are quite different, reflecting distinct habits for each country, as discussed previously.

The curves for weekends have very distinct peaks of activities from those of weekdays, both at the country and city levels. For instance, as shown in Figure \ref{fig:chkCountryWeekdayWeekend}, English people have a very distinct drinking pattern from Americans on weekends. Moreover, the differences among the countries in terms of preferences at Slow Food places are also clear on weekends: Brazilians tend to go to Slow Food places more often at lunch time, whereas Americans and English people do it more at dinner time.

We note that there is no clear (dominant) temporal \checkin pattern for Fast Food places on weekends, when considering different cities of a country. However, we do note that most activities happen after noon, which was expected. In contrast, there is a dominant pattern for \checkins at Slow Food places on the weekends, and it is similar to the one observed on weekdays. This is possibly because such places (often restaurants) have well-defined opening hours, serving meals around lunch and dinner times only, which coincide with the times of \checkin peaks (Figures \ref{fig:chkCountryWeekdayWeekend}c, \ref{fig:chkCountryWeekdayWeekend}f, \ref{fig:chkCityWeekdayWeekend}c, and \ref{fig:chkCityWeekdayWeekend}f). Assuming that the height of such peaks reflects the importance of that meal for a certain culture, we note once again a key distinction between Americans and Brazilians.

\subsection{Discussion}

In addition to temporal and spatial patterns of \checkins at different types of places, we also compute the Shannon's entropy~\cite{entropyShannon} of preferences for each venue subcategory among all considered areas. The goal is to analyze whether the \checkins at specific subcategories are more concentrated at specific areas (low entropy) or not (high entropy). We compute the entropy for subcategories of each class (Drink, Fast Food and Slow Food) at country and city levels. The average entropy for subcategories of the Drink class is 3.23 (standard deviation $\sigma=0.93$) for countries and is 3.88 ($\sigma=1.09$) for cities. Sake bar is one example with low entropy (1.13 for countries and 1.89 for cities), which indicates that this subcategory is popular on very few countries and cities. Surely Japan contributes considerably to this result. On the other hand, the average entropy for subcategories of the Slow Food class is much larger, 2.63 ($\sigma=0.78$). This higher entropy reflects the widespread 
popularization of various cuisines. For example, a \checkin at an Italian restaurant does not necessarily mean that it represents a behavior of an Italian, since it is a very international type of restaurant, confirmed by the high entropy (3.63). Note, however, that if the \checkin at an Italian restaurant is made at lunch time it could be more likely to represent a Brazilian behavior than American, since Brazilians have their main meal at lunch time, as presented in Section~\ref{secTemporalAna}. Time plays an important role in this case.

Given these considerations and all the observations reported here, we propose the use of spatio-temporal correlations of \checkins as cultural signatures of regions.

\section{Identifying Cultural Boundaries}\label{secCulturalBound}

\subsection{Clustering Regions}

In this section, we use the cultural signatures of regions described above to identify similar areas around the planet according to their cultural aspects, delineating their so-called ``cultural boundaries''. To that end, we first represent each area $a$ by a high dimensional preference vector composed of 808 features, namely the normalized number of \checkins at each of the 101 subcategories in four disjoint periods of the day, on weekdays and on the weekends. We then apply the Principal Component Analysis (PCA)~\cite{pca} technique to these vectors to obtain their principal components\footnote{Alternative methods could be applied to reduce the dimensionality of these vectors. A comparison of these methods is out of the scope of the present work.}. Finally, we use the $k$-means algorithm, a widely used clustering technique, to group areas in the space defined by these principal components. We perform this analysis for areas defined at the country, city and neighborhood levels.

\begin{figure*}[t!]
\centering
\subfigure[Countries]
%{\includegraphics[width=.27\textwidth]{../images/drink/dadosCountries/distancias/pca/countries-all.pdf}}
{\includegraphics[width=.27\textwidth]{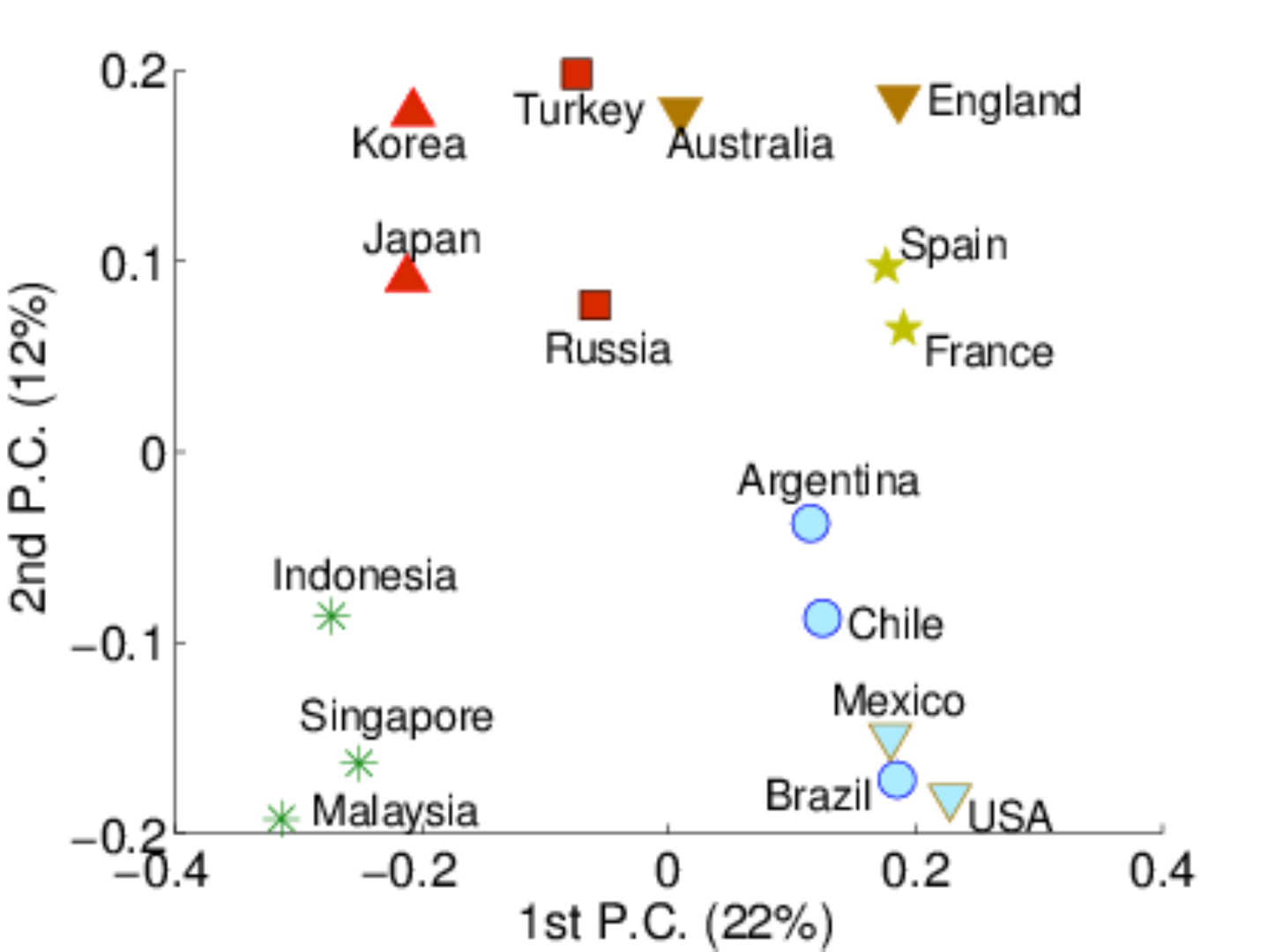}}
\subfigure[Cities]
%{\includegraphics[width=.27\textwidth]{../images/drink/dadosCities/zoomOut_6x/clusters/pca/cities-all.pdf}}
{\includegraphics[width=.27\textwidth]{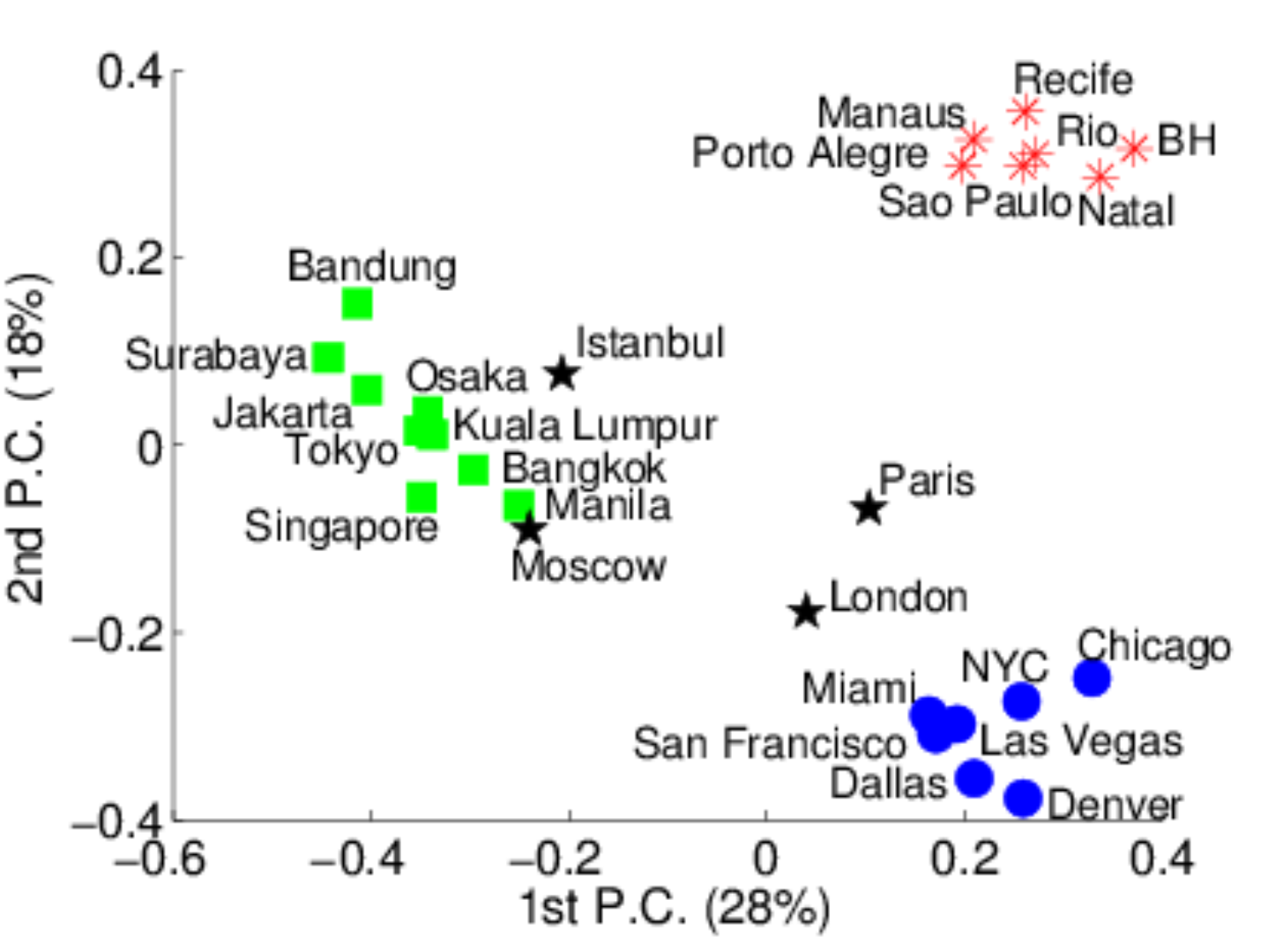}}
\subfigure[Regions]
%{\includegraphics[width=.27\textwidth]{../images/drink/regions_city/NY_LONDON_TOKYO/cluster/pca/regions-all.pdf}}\vspace{-4mm}
{\includegraphics[width=.27\textwidth]{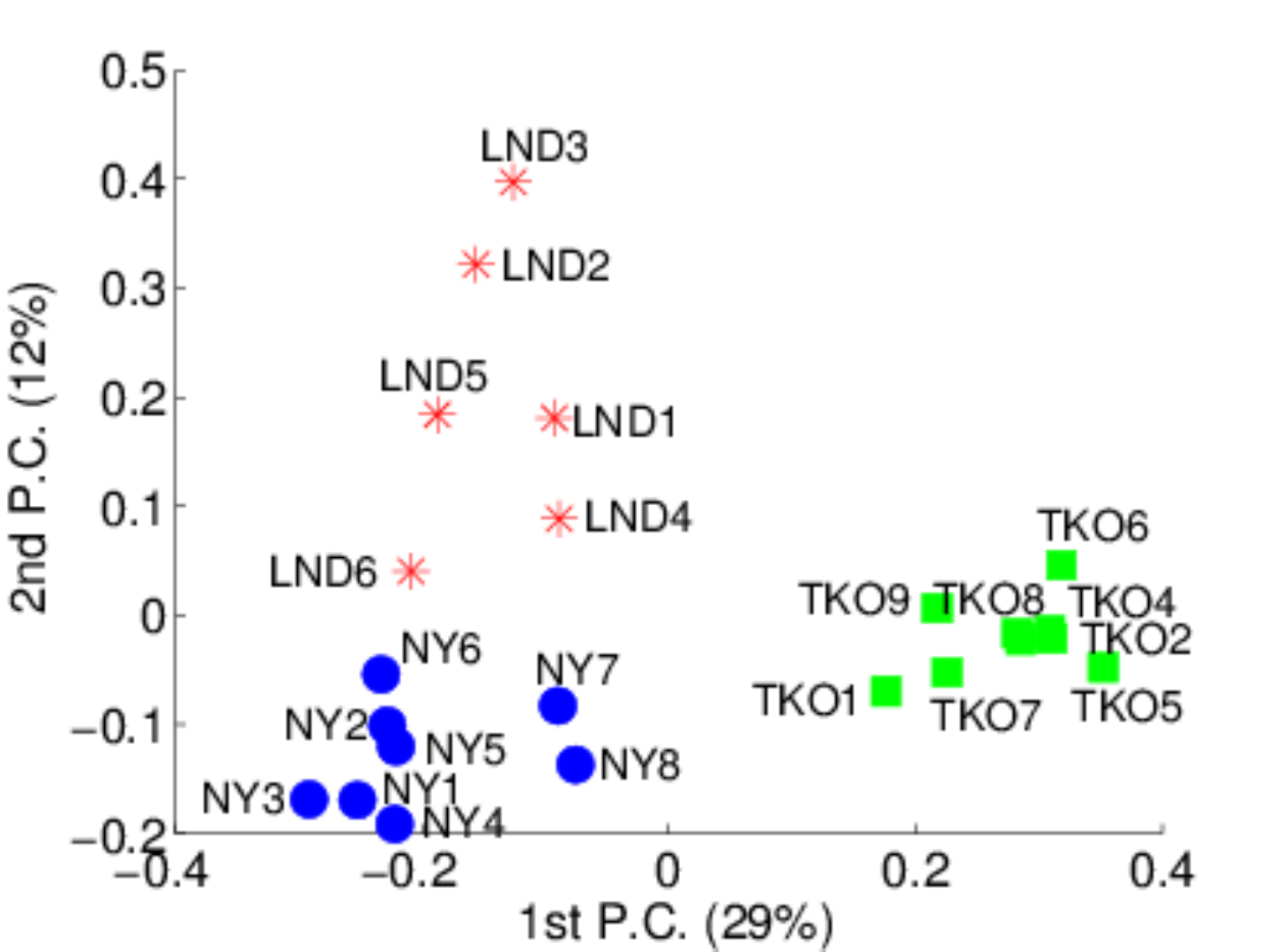}}
\caption{Clustering results for countries, cities, and regions inside cities.}\label{fig:pcaCountriesCities}
\end{figure*}

The score values for the first two principal components generated by the PCA for countries, cities, and regions are shown in Figures~\ref{fig:pcaCountriesCities}a, \ref{fig:pcaCountriesCities}b, and \ref{fig:pcaCountriesCities}c, respectively. The variance in the data explained by these first two components is shown in each figure. Each color/symbol in those figures indicates a cluster obtained by $k$-means, which used the $p$ first principal components that explain 100\% of the variation in the data ($p$=$15$ for countries, $p$=$26$ for cities and $p$=$22$ for regions). The $k$ value in the $k$-means varied according to the characteristics of the considered areas. For countries, we set $k$=$7$ (same number of clusters used in~\cite{inglehart:2010}). Following the same logic, we set $k$=$4$ for cities, since we considered cities from 4 different continents/countries, and $k$=$3$ for regions inside a city, because we considered 3 cities. We used the cosine similarity to 
compute the similarity between locations.

It is possible to observe in Figure~\ref{fig:pcaCountriesCities}a that countries with closer geographical proximity are not necessarily associated with the same cluster. For example, Australia and Indonesia are {\it not} in the same cluster. Although they are geographically neighboring countries, they are culturally very distinct. When analyzing large cities from the considered countries, Figure~\ref{fig:pcaCountriesCities}b shows that they are well clustered by the geographical regions where they are located: Asia, Brazil, Europe and USA. Intuitively, this result makes sense, since, for instance, cosmopolitan European capitals tend to present more similar cultural habits among each other than among cities from different continents. Turning our attention to regions inside London, NY, and Tokyo, we observe in Figure~\ref{fig:pcaCountriesCities}c that all regions in the same city are in the same cluster. This result was also expected when considering all features. Besides that, when we analyze a subset of 
features, for example, drinking habits during weekends in all regions of London, NY, and Tokyo (result omitted), we find that some regions of London and NY are clustered together. This is corroborated by the results shown in Section~\ref{sec:featuresExtract}: for certain categories, there are regions from different cities that are very similar and, thus, end up clustered together.

\subsection{Comparing with Survey Data}

\begin{figure}[t!]
\centering
%\subfigure[Given by the WVS]
{\includegraphics[width=.30\textwidth]{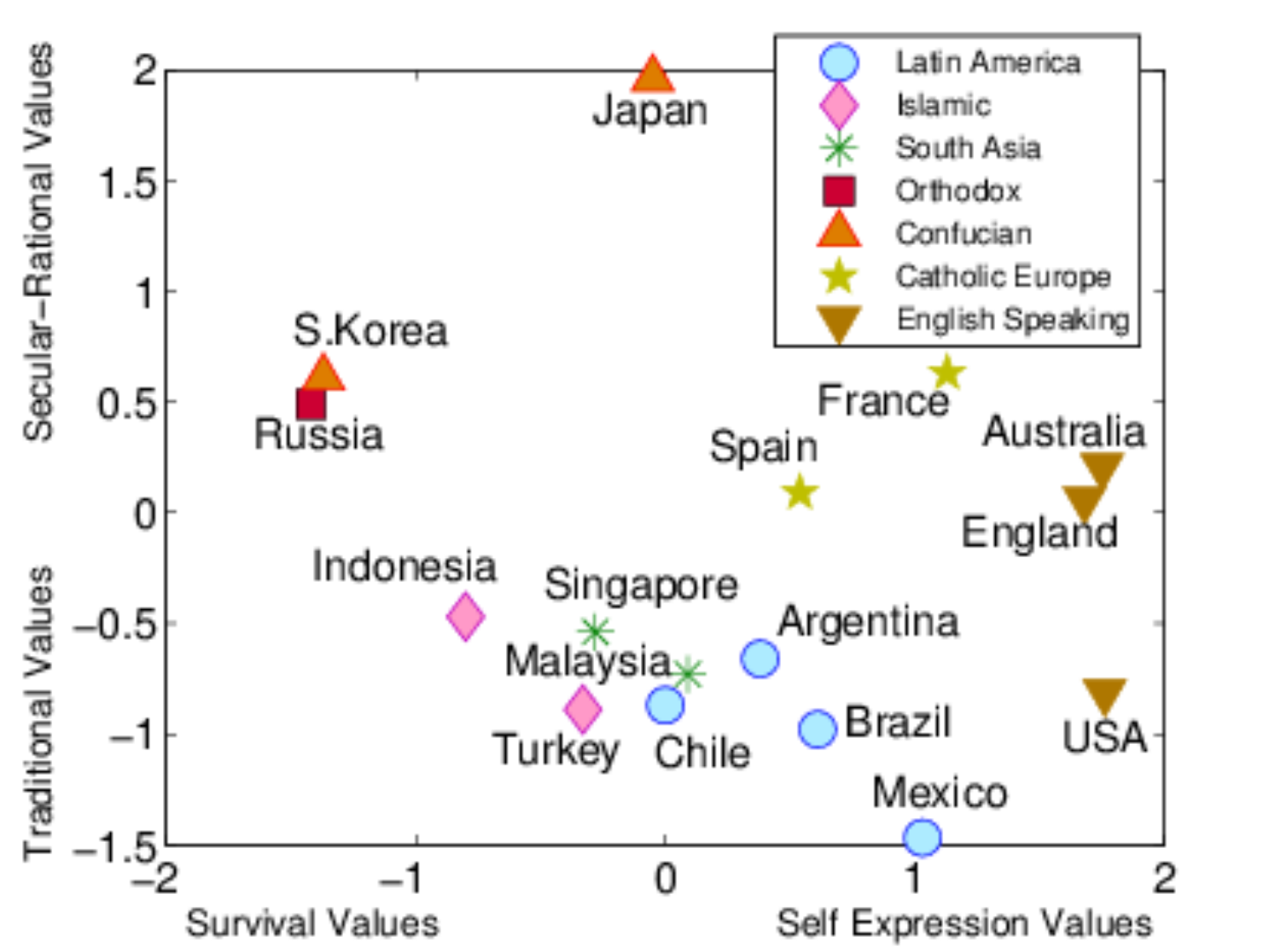}}%\vspace{-2.3mm}
%\subfigure[Given by our approach.]
%{\includegraphics[width=.40\textwidth]{../images/pca/pcaCountriesAll-k7.pdf}}
\caption{The cultural map of the World given by the World Values Survey.}\label{fig:wvs}
\end{figure}

Similarly to us, Ronald Inglehart and Christian Welzel proposed a cultural map of the world based on the World Values Surveys (WVS) data from 2005 to 2008~\cite{inglehart:2010}. This map is shown in Figure~\ref{fig:wvs} and contains only the countries we analyze in this paper. It reveals two major dimensions of cross-cultural variation: a traditional versus secular-rational values dimension and a survival versus self-expression values dimension. Moreover, it offers a division of the world into clusters, similarly to what we have done in the previous section. Comparing Figures~\ref{fig:pcaCountriesCities}a~and~\ref{fig:wvs}, observe that the similarities are striking, with only two major differences. First, the ``Islamic'' cluster dissolved, with Turkey joining Russia and Indonesia joining Malaysia and Singapore. Second, USA and Mexico left the ``English Speaking'' and the ``Latin America'' clusters, respectively, and paired up to form a new one. Note, nevertheless, that these differences might not be 
surprising as these new boundaries.

%Note, nevertheless, that these differences are very reasonable and it would not be surprising to us if one argues that these new boundaries agree more with the common sense than those extracted from the WVS.

We formally investigate the differences between boundaries given by the WVS study and by our approach. In order to do so we rank, for a given country, all the other countries according to  their cosine similarity towards it. We compute the similarity using the dimensions produced by the WVS data~\cite{inglehart:2010} and the dimensions computed by our approach. Then, we compute the Spearman's rank correlation coefficient $\rho$ between these two ranks to see, for instance, if the most similar (and distinct) countries to England using the WVS data are ranked similarly when we use our approach. In our approach, we use two different datasets. In $dataset_1$, we use the full set of features, as done so far. In $dataset_2$, we use solely the features extracted from the fast food \checkins performed during the weekends\footnote{This particular set of features was chosen because it was the configuration which gave the best results.}. Table 1 shows these results. We highlight in bold all the 
coefficients which are statistically significant, i.e., with a $p$-$value < 0.05$. Observe that the correlation $\rho$ is significant and positive for several countries. For $dataset_1$ and $dataset_2$, 9 and 12 
countries have similar ranks with the ones given by the WVS, respectively. This shows that our approach, which is based solely on one week of participatory data, has a clear potential to reproduce cultural studies performed using surveys, such as the ones relying on the WVS, which is based on 4 years of survey data.

\begin{table}[t!]
\caption{The Spearman's rank correlation coefficient $\rho$ (and its respective p-value) between the rank of similar countries generated from WVS and by our approach.} 
\begin{center}
\tiny
    \begin{tabular}{ | c | c | c | c | c |}
    \hline
		Country & \multicolumn{2}{|c|}{$dataset_1$} & \multicolumn{2}{|c|}{$dataset_2$} \\ \hline
    - & $\rho$ & p-value & $\rho$ & p-value \\ \hline
		Argentina &\textbf{0.56} &\textbf{0.03} &\textbf{0.77} &\textbf{0.0007}  \\ \hline
		Australia &0.32 &0.23 &\textbf{0.60} &\textbf{0.02} \\ \hline
		Brazil &0.48 &0.06 &\textbf{0.81} &\textbf{0.0002} \\ \hline
		Chile &0.32 &0.23 &\textbf{0.53} &\textbf{0.04} \\ \hline
		England &\textbf{0.87} &\textbf{0} &\textbf{0.70} &\textbf{0.004} \\ \hline
		France &\textbf{0.85} &\textbf{2e-06} &\textbf{0.61} &\textbf{0.01} \\ \hline
		Indonesia &\textbf{0.84} &\textbf{4e-05} &\textbf{0.75} &\textbf{0.001} \\ \hline
		Japan &0.38 &0.15 &0.39 &0.13 \\ \hline
		Korea &\textbf{0.68} &\textbf{0.004} &0.45 &0.08 \\ \hline
		Malaysia &-0.16 &0.54 &0.11 &0.68 \\ \hline
		Mexico &\textbf{0.55} &\textbf{0.03} &\textbf{0.71} &\textbf{0.003} \\ \hline
		Russia &\textbf{0.78} &\textbf{0.0006} &\textbf{0.76} &\textbf{0.001} \\ \hline
		Singapore &0.34 &0.20 &\textbf{0.65} &\textbf{0.008} \\ \hline
		Spain &\textbf{0.78} &\textbf{0.0005} &\textbf{0.75} &\textbf{0.001} \\ \hline
		Turkey &-0.18 &0.50 &-0.31 &0.24 \\ \hline
		USA &\textbf{0.70} &\textbf{0.004} &\textbf{0.67} &\textbf{0.005} \\ \hline
    \end{tabular}
\end{center}
\end{table}
\label{tab:cmpsurvey}
\normalsize

We would also like to point out the reasons for the differences between our cultural map and the WVS map, as well as for the negative correlations seen in Table 1. First, the traits of each dataset are significantly different. While the WVS looked at several cultural dimensions, from religion to politics, from economics to lifestyle, we looked only at food and drink preferences. Second, the WVS data has a distance of $4$ to $7$ years to our data. During this time, significant cultural changes may have happened, given that the world is getting more connected at every day. Third, the most significant differences are related to  multi-ethnic, multicultural, and multilingual countries, such as Malaysia and Turkey. In these countries it is probably hard to find culturally homogeneous samples of individuals, which might be the cause of the discrepancies seen between our results and those described in~\cite{inglehart:2010}.

\section{Conclusions and Future Work}\label{sec:conclusion}

This work proposes a new methodology for identifying cultural boundaries and similarities across populations. For that, we map food and drink \checkins extracted from Foursquare into users' cultural preferences, considering spatio-temporal dimensions. We then apply a simple clustering technique to show the ``cultural distance'' among countries, cities or even regions within a city. The considered set of features allows the identification of cultural boundaries that despite often agreeing on common knowledge, is based on large-scale data. Thus, unlike other empirical work, which is based on survey data, our methodology can reach global scale much faster and at a much lower cost. It is also important to emphasize that the proposed methodology could be used to work with other types of features, which might be useful for other kind of studies. 

One of the obvious directions is to exploit the criteria for identifying cultural boundaries defined in this paper in order to perform social studies at large scale. Besides that, we also want to develop recommendation mechanisms considering the cultural characterization of specific urban areas. This could be useful, for instance, for location-based social networks like Foursquare to improve their current recommendation systems. 

\section*{Acknowledgments}
\scriptsize{This work is partially supported by the INCT-Web (MCT/CNPq grant 57.3871/2008-6), and by the authors’ individual grants and scholarships from CNPq, CAPES, FAPEMIG, and EPSRC Grant ``The Uncertainty of Identity'' (EP/J005266/1).}

\footnotesize{
\bibliographystyle{aaai}

}

\end{document}